\documentclass[amsmath,amssymb,aps,prl,reprint,floatfix,superscriptaddress]{revtex4-1}

\usepackage{graphicx}
\usepackage{subfigure}
\usepackage{dcolumn}
\usepackage{bm}
\usepackage[none]{hyphenat}
\usepackage[mathlines]{lineno}
\usepackage{float}
\usepackage[utf8]{inputenc}
\usepackage [english]{babel}
\usepackage [autostyle, english = american]{csquotes}
\MakeOuterQuote{"}
\usepackage{units}

\newcommand{\ket}[1]{\left|#1\right\rangle}

\newcommand{\braket}[2]{\left\langle #1|#2\right\rangle}
\newcommand{\ord}[1]{\mathcal{O}\left(#1\right)}

\newcommand{\rom}[1]{\uppercase\expandafter{\romannumeral #1\relax}}

\begin{document}

\title{Quantum simulations with complex geometries and synthetic gauge fields in a trapped ion chain}


\author{Tom Manovitz}
\email{tom.manovitz@weizmann.ac.il} \thanks{Equal contribution}
\author{Yotam Shapira} \thanks{Equal contribution}
\author{Nitzan Akerman}
\affiliation{%
Department of Physics of Complex Systems, Weizmann Institute of Science, Rehovot 7610001, Israel
}%
\author{Ady Stern}
\affiliation{%
Department of Condensed Matter Physics, Weizmann Institute of Science, Rehovot 7610001, Israel
}%
\author{Roee Ozeri}
\affiliation{%
Department of Physics of Complex Systems, Weizmann Institute of Science, Rehovot 7610001, Israel
}%

\begin{abstract}
In recent years, arrays of atomic ions in a linear RF trap have proven to be a particularly successful platform for quantum simulation. However, a wide range of quantum models and phenomena have, so far, remained beyond the reach of such simulators. In this work we introduce a technique that can substantially extend this reach using an external field gradient along the ion chain and a global, uniform driving field. The technique can be used to generate both static and time-varying synthetic gauge fields in a linear chain of trapped ions, and enables continuous simulation of a variety of coupling geometries and topologies, including periodic boundary conditions and high dimensional Hamiltonians. We describe the technique, derive the corresponding effective Hamiltonian, propose a number of variations, and discuss the possibility of scaling to quantum-advantage sized simulators. Additionally, we suggest several possible implementations and briefly examine two: the Aharonov-Bohm ring and the frustrated triangular ladder.
\end{abstract}

\maketitle

\section{Introduction}

Quantum simulators are highly controlled quantum machines with which it is possible to engineer and study complex quantum states and dynamics. Such machines, when large and accurate enough, are expected to elucidate the behaviour of quantum systems that defy analytical treatment and which are intractable for classical numerical simulations \cite{feynman1982simulating}. With the increasing sizes and abilities of quantum simulators and computers \cite{bernien2017probing,zhang2017observation,arute2019quantum,Bohnet2016}, quantum advantage in the context of quantum simulation may be within reach in the near future \cite{Acin2018,bermejo2018architectures}. Of the diverse physical platforms used for quantum simulation, atomic ion chains in linear RF traps have proven particularly fertile by virtue of their long coherence times and high operation fidelity \cite{Blatt2012,monroe2019programmable}. Using trapped ion quantum simulators, researchers have created and studied a wealth of quantum phenomena by applying both analog \cite{gerritsma2010quantum,Kim2010,Islam2011,islam2013emergence,schachenmayer2013,Richerme2014,Jurcevic2014,Smith2016,jurcevic2017,zhang2017observation,zhang2017discrete,Zhang2018,Gorman2018} and digital  \cite{Barreiro2011, lanyon2011universal, Schindler2013, Martinez2016, Hempel2018} simulation techniques. 

A principal feature of ion chain quantum simulators is the precisely controllable long-range coupling between ion-qubits, driven by an external field and mediated by the motional modes of the chain \cite{molmer1999multiparticle,sorensen2000entanglement,roos2008}. Representing a spin-$\frac{1}{2}$ particle by two electronic energy levels in each ion, a uniform external driving field can induce an effective spin-spin interaction of the form \cite{porras2004effective}:
\begin{equation} \label{spinspin}
    H_c=\sum_{i<j,\alpha}J_{ij}^\alpha\sigma^\alpha_i\sigma^\alpha_j
\end{equation}
where $i,j$ denote the spin index, $\sigma^\alpha_i$ with $\alpha\in\{x,y,z\}$ are the standard Pauli operators acting on spin $i$, and $J^\alpha_{ij}$ represents the coupling matrix for the different Pauli axes. Often, quantum simulation experiments with trapped ions use a uniform bichromatic field to couple the spins through the transversal motional modes of the ion chain, generating a coupling matrix \cite{porras2004effective,islam2013emergence,monroe2019programmable}:
\begin{equation} \label{jij}
    J_{ij}\sim\frac{1}{|i-j|^q},\quad 0<q<3.
\end{equation}

Despite the effectiveness of these tools, many territories remain uncharted for linear ion trap quantum simulators. One outstanding challenge is that of simulating systems in more than a single spatial dimension.  The successes of 1D ion trap quantum simulators calls for extending their scope to explore the richness of higher dimensional quantum systems. However, ion chains are open-ended and one dimensional, and the couplings that are induced by the simplest and most robust simulation techniques, expressed in Eq. \eqref{jij}, naturally reflect this geometry.  
In the past several years, new methods have been developed in order to enable more complex coupling geometries \cite{Korenblit2012,rajabi2019dynamical,shapira2019theory,davoudi2019towards,lu2019global,figgatt2019parallel}. Despite these advancements, hardly any scalable simulations of high dimensional Hamiltonians have been shown in an ion chain.

Another tool that linear ion trap simulators currently lack, yet may aspire to, is the simulation of magnetic fluxes. Magnetic fluxes are a key ingredient in a range of quantum phenomena, with the iconic example being the quantum Hall effect \cite{hofstader1976butterfly,tknn1982iqhe}. Such fluxes can be expressed in the Hamiltonian by complex coupling terms, such as $e^{i\phi}\psi_i^\dagger\psi_j$, representing the presence of a gauge field potential which associates a phase to a directional propagation of an excitation along the lattice, also known as a Peierls phase. For this reason, the synthetic or artificial generation of gauge field interaction terms has emerged in recent years as one of the most prolific tools of neutral atom quantum simulation \cite{jaksch2003creation,miyake2013realizing,dalibard2011colloquium,bloch2012quantum,lin2011spin,struck2013engineering,lin2009synthetic,aidelsburger2013realization,mancini2015observation}, and might similarly present new opportunities in trapped ion quantum simulators. 

In principle, making use of the universal gate set already available in ion trap quantum computers \cite{gaebler2016,debnath2016demonstration,lanyon2011universal}, one can digitally simulate any quantum system by breaking down the dynamics into a series of simpler operations \cite{lanyon2011universal,Blatt2012,barenco1995}; such a simulation can include all features discussed above. However, in practice, the engineering cost of a universal gate set is high and the decomposition of target models may be unwieldy and can incur a high fidelity cost. While in the long term digital simulations may benefit from fault-tolerant quantum error correction, the necessary qubit array sizes and operation fidelities to reach this threshold are far beyond current capabilities. Hence, analog quantum simulations, in which the target Hamiltonian is continuously implemented, arguably offer more promising prospects for near and mid-term quantum simulation \cite{preskill2018quantum}. This motivates expanding the range of models that are directly simulatable with trapped ions.

In this manuscript we introduce a scalable and experimentally simple technique that can be used to simulate a large range of spin Hamiltonians on an ion chain. This technique improves on the standard schemes in two significant ways: through generation of complex coupling geometries, including high dimensional Hamiltonians and closed boundary conditions; and by an introduction of both a static and time-dependent Peierls phase, effectively generating a synthetic gauge field. Crucially, the technique can be performed with a uniform intensity global driving beam, with no dynamical control, and with fields that are independent of the number of ions. The technique requires an addition of an external gradient field along the chain and the use of a multitone driving field.

Gradient fields have been most prominently used in ion chains in quantum processing architectures where a strong spectral separation enables both individual addressing of the ions (as in NMR) as well as driving entangling gates using long-wavelength fields \cite{mintert2001ion,johanning2009individual,timoney2011quantum}. The technique outlined in this manuscript can be understood as an extension or variation of the NMR-inspired scheme, spectrally resolving coupling terms rather than individual subsystems. A similar proposal for employing gradient fields to simulate higher dimensional systems was recently put forth by Rajabi et al. \cite{rajabi2019dynamical}, requiring the additional use of dynamical techniques. 

We note exciting proposals for simulation of static \cite{grass2015,grass2018} and dynamic \cite{hauke2013quantum,davoudi2019towards} gauge fields as well as high dimensional Hamiltonians \cite{Korenblit2012,davoudi2019towards} in a trapped ion chain using either dynamical techniques \cite{grass2015,grass2018}, individual addressing of all ions \cite{Korenblit2012,davoudi2019towards}, or additional energy levels \cite{hauke2013quantum}. Furthermore, proposals have been put forth for implementing synthetic gauge fields for the motional, rather than electronic, degrees of freedom of the ion chain \cite{bermudez2011synthetic,bermudez2012photon,vermersch2016implementation}. 

The manuscript is ordered as follows: we first present our main result; we then describe the technique's principle of operation and derive the coupling Hamiltonian; we use the Aharonov-Bohm ring in order to exemplify the salient features of our technique; we suggest a number of simulatable Hamiltonians of interest; and finally, we discuss the challenges posed for realizing the technique on large quantum simulators. In the appendix we describe variations that can ease implementation and further increase the range of target models.

\section{Main results}

\begin{figure}
    \includegraphics[width=\columnwidth]{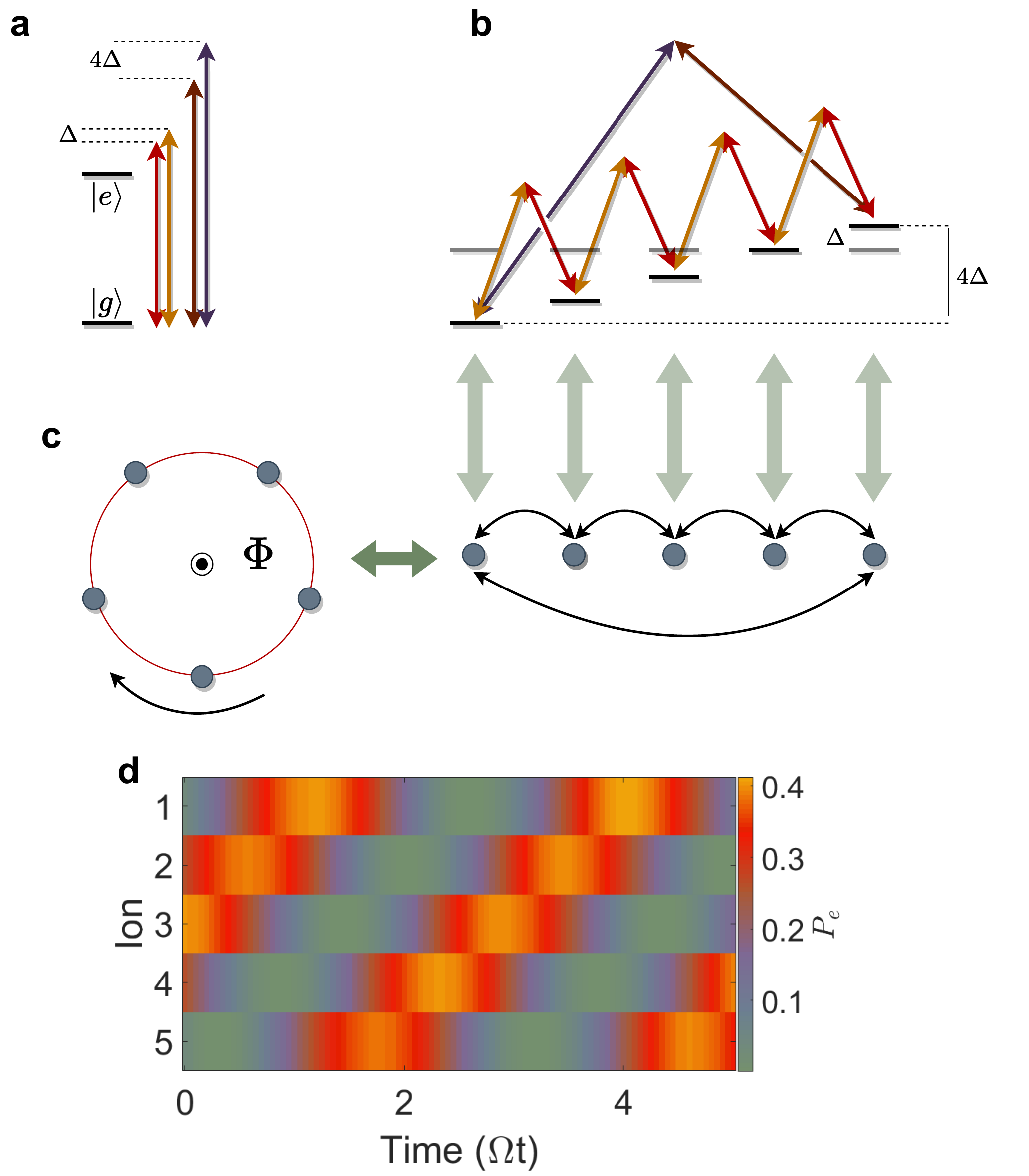}
	\caption{Quantum simulation with synthetic gauge fields and complex geometries. Driving field pairs (a) bridge the resonance difference between ions created by an external gradient (b). Here we only plot a pair of blue sidebands for illustration purposes, where the corresponding red sideband pair is not shown.  The controlled resonances can be used to tailor different coupling geometries, such as 1D rings (c). The phases of the driving pairs can be chosen to generate synthetic gauge fields, representing magnetic fluxes threading the lattice. In this example, as a result, an excitation on the lattice will be driven around the ring, in a direction and velocity dictated by the flux (d). Here the flux is $\Phi=3\pi/4$. The probability $P_e$ for a local spin excitation is color coded.
	}
	\label{mainresult}
\end{figure}

Our main result is a simple recipe for generating a class of Hamiltonians using trapped ions. The class is described by the following formula:
\begin{equation} \label{genform}
    H=\sum_{n=1}^{N-1}H_n=\sum_{n=1}^{N-1}\Omega_n e^{i(\phi_n-\delta_nt)}\sum_{i=1}^{N-n}\sigma^{+}_i\sigma^{-}_{i+n} + h.c.
\end{equation}
where $\sigma_i^+$ $(\sigma_i^-)$ denotes the raising (lowering) Pauli operator on ion $i$, and $\Omega_n$, $\phi_n$ and $\delta_n$ are tunable parameters corresponding to the coupling strengths, static phases and time-dependent phases, respectively, of an $n$-neighbor hopping interaction. As we show below, the Hamiltonian in Eq. \eqref{genform} can be used to implement spin Hamiltonians on various geometries, and can furthermore manifest static and time-dependent gauge fields.

The method relies on the application  of a static external field gradient (e.g. a spatially varying magnetic or light shift field) for shifting the atomic energy levels along the ion chain, applied together with  a corresponding global uniform driving field. The field gradient collapses the translational symmetry of the ion chain, effectively suppressing the standard coupling form of Eq. \eqref{jij}. However, because the gradient is  spatially uniform, the driving field can be used to selectively reinstate the translational symmetry of the spin-spin interaction. This is done in a controlled manner by bridging the resonance difference between equally-separated ion-pairs using the frequency difference between pairs of bichromatic fields. Furthermore, the breaking of spatial symmetry differentiates the interaction of an ion with its neighbors to the left and right, which gives rise to a gauge-field-like phase and an effective breaking of time-reversal symmetry.

The tools needed to implement (\ref{genform}) are standard in trapped ion experiments: for every nonzero $\Omega_n$, representing an $n$-neighbor interaction, a four-tone field is added. The corresponding $\Omega_n$, $\phi_n$ and $\delta_n$ are set by the field's amplitudes, phases and frequencies. The driving field is activated using a single uniform-intensity beam. The magnetic field gradient can be modest, on the order of $10$ G/cm. Additionally, the interaction in Eq. \eqref{genform} is excitation-number-preserving implying robustness to global dephasing noise.

A wide spectrum of quantum phenomena can be accessed using this method. For example, by choosing $\Omega_1=\Omega_{N-1}=\Omega$ and $\phi_1=-\phi_{N-1}=2\pi\Phi/N$ (and nulling all other parameters) we arrive at a lattice ring Hamiltonian with a hopping term:
\begin{equation} \label{ring}
    H_r\left(\Phi\right)=\Omega\sum_{i=1}^N e^{2\pi i\Phi/N}\sigma^+_i\sigma^-_{i+1}+h.c.
\end{equation}
where boundary conditions are periodic. The phase $\Phi$ corresponds to the Aharonov-Bohm phase acquired by an electric charge encircling a ring penetrated by a magnetic flux (note that a 1D spin system may always be described as a fermionic system, using the Jordan-Wigner transformation) \cite{aharonov1959}. Accordingly, an excitation will travel clockwise or counter-clockwise on the ring, generating a persistent current \cite{roushan2017chiral}. While this model is easy to solve, it clearly showcases the main tools of the proposed technique. We analyze the model in more detail below. 

Figure (\ref{mainresult}) highlights our method's main principles of operation. Taking the 1D ring in a $N=5$ ion chain as an example, a pair of driving fields (a) bridge the energy difference between the ion created by the external gradient, i.e $\Delta$ for neighboring ions and $4\Delta$ for the edge ions (b). These driving fields form tailored couplings between the ions, and are here used to generate a 5-site ring penetrated by a magnetic flux $\Phi$ (c). Accordingly, a simulation of the ion chain's evolution shows an excitation travelling around the ring (d).

\begin{figure}
 \includegraphics[width=\columnwidth]{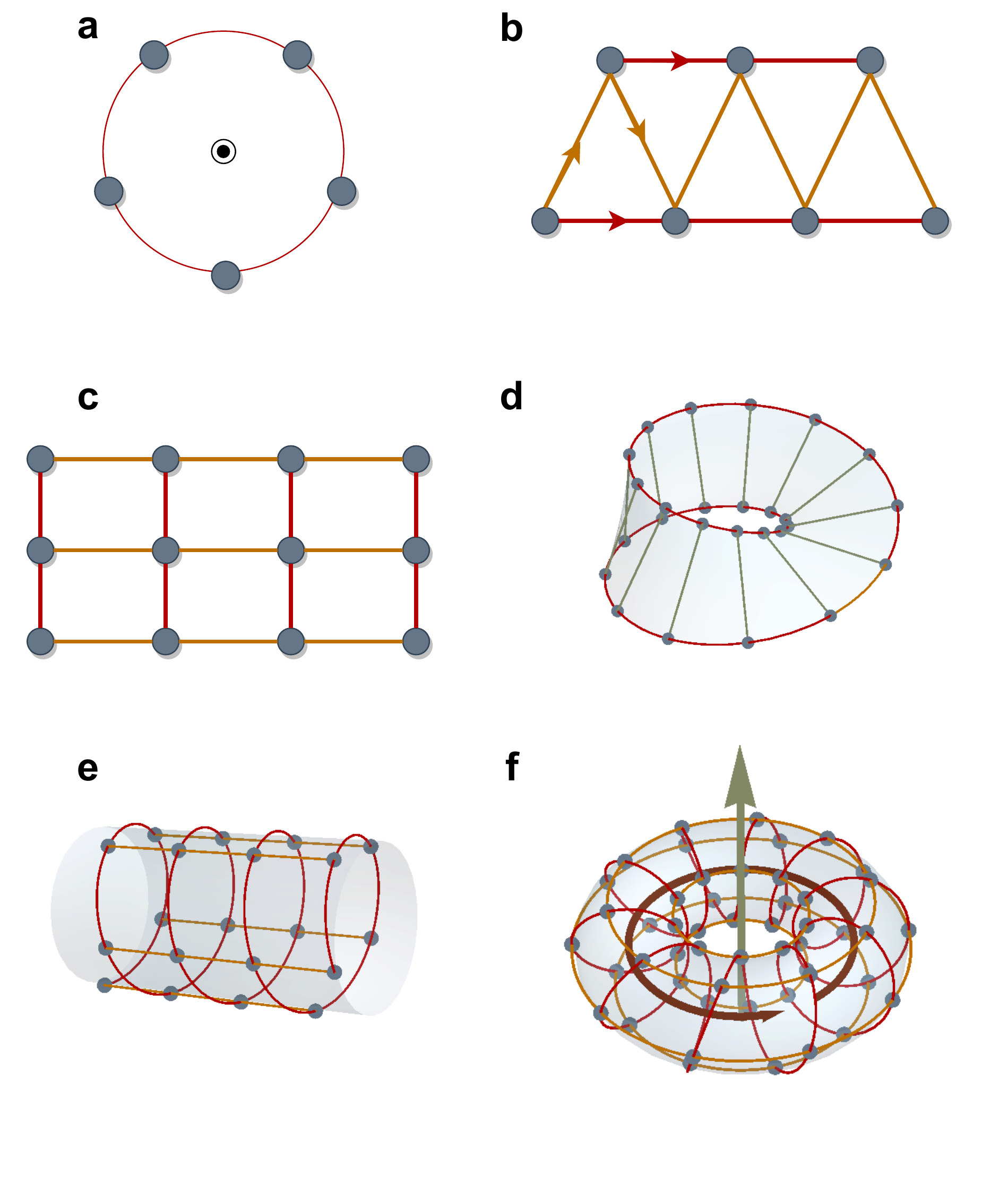}
\caption{Implementation of various geometries using appropriate hopping interactions. (a) nearest-neighbour and $N-1$ neighbor interactions generate a ring; (b) nearest-neighbour and next-nearest-neighbour interactions generate a triangular ladder; (c) nearest-neighbour and $W$-neighbor interactions generate a rectangular lattice when used with spacer ions (see discussion below), and with the addition of $N/W$ interactions generate a rectangular lattice on a cylinder; (d) nearest-neighbour, $N/2$ and $N-1$ interactions generate a M{\"o}bius ladder; (e) nearest-neighbour and $W$ interactions generate a helical lattice on a cylinder; (f) by adding $N-1$ and $N/W$ interactions to the helical lattice, the cylinder is closed onto a torus. Controlling the phases of these interactions results in synthetic gauge fields representing fluxes threading these geometries. For instance, in the torus (f) an external axial flux (green), or within the torus, as in an anapole moment (red), can be produced.}
\label{models}
\end{figure}

Using these principles, Hamiltonians of the form expressed in Eq. \eqref{genform} can be generated. Significantly, a wide variety of coupling geometries are reachable. We illustrate some possible coupling geometries in Figure \ref{models}. Besides a ring (a), these include triangular ladders (b); 2d rectangular lattices (c) (which can be closed onto a cylinder, not shown); a M{\"o}bius-strip ladder (d); a helical lattice on a cylinder (e); and a torus (f). These lattices can be threaded by a variety of magnetic fluxes, as illustrated for the torus (f). While the 1D ring can be mapped to a system of free noninteracting fermions and is thus simply solvable, all other models shown here are expected to show complex behavior and can be difficult to solve. 

\section{Physical picture}

Ions in a Paul trap are frequently modeled as two level spins with a set of harmonic modes, where the former corresponds to the ions' electronic degrees of freedom, and the latter to the motional normal-modes of the ion chain. External electromagnetic fields can couple to spin and motional degrees of freedom and, with proper tuning, can be used to engineer effective interactions between the spins of different ions via mediation by the motional modes.

In the M{\o}lmer-S{\o}rensen (MS) interaction \cite{molmer1999multiparticle,sorensen2000entanglement}, the external field is bichromatic and tuned to frequencies $\omega_\pm=\omega_0\pm\left(\nu+\xi\right)$, with $\hbar\omega_0$ the single qubit energy separation, $\nu$ the frequency of a normal mode of motion of the ion-chain, and $\xi$ a constant detuning which together with the field intensity determines the interaction rate. The tone $\omega_+$ ($\omega_-$) mediates interactions via the blue (red) motional sideband, i.e it employs transitions which excite the ion's electronic degree of freedom while adding (removing) a phonon of the motional normal-mode. While only two driving tones are used, this interaction couples any two ions in the chain in four different "pathways", as is shown in Fig. 1 of Ref. \cite{sorensen2000entanglement}. 

When coupled to the center-of-mass (COM) mode of the ion chain, this driving field induces an effective $\sigma^x\sigma^x$ interaction through a two photon process, which equally couples all of the ions-pairs in the ion chain. This interaction can be decomposed to two contributions: a pair creation/annihilation term, $\sigma^+\sigma^++h.c$, which drives a $\ket{\uparrow\uparrow}\leftrightarrow\ket{\downarrow\downarrow}$  two photon transition, changing the system energy by $\omega_+ + \omega_- = 2\omega_0$; and an excitation hopping term $\sigma^+\sigma^-+h.c$, driving a $\ket{\downarrow\uparrow}\leftrightarrow\ket{\uparrow\downarrow}$  two photon transition, which leaves the system's energy unchanged, i.e $\omega_\pm - \omega_\pm=0$. 

Here we are interested in eliminating the pair creation/annihilation term while retaining the hopping term, and furthermore shaping its coupling matrix. The first goal is achieved by detuning the driving field frequencies from resonance with the two-photon transition, i.e by modifying the bichromatic drive frequencies to $\omega_\pm=\omega_0+\epsilon\pm\left(\nu+\xi\right)$, shifting the pair creation/annihilation term $2\epsilon$ off-resonance.
The second goal requires a more elaborate approach. As the $\sigma^+\sigma^-+h.c$ term is mediated by an excitation/de-excitation pair of identical photons, it resonantly couples only states that are degenerate under $H_0$; this implies a coupling between all equal-excitation states. However, the hopping term can also be controllably suppressed by lifting the equal-excitation degeneracy \cite{shaniv2018toward}. A controlled suppression of the interaction will then allow for selectively reinstating resonant conditions through a modulation of the driving field.

\begin{figure}
    \includegraphics[width=\columnwidth]{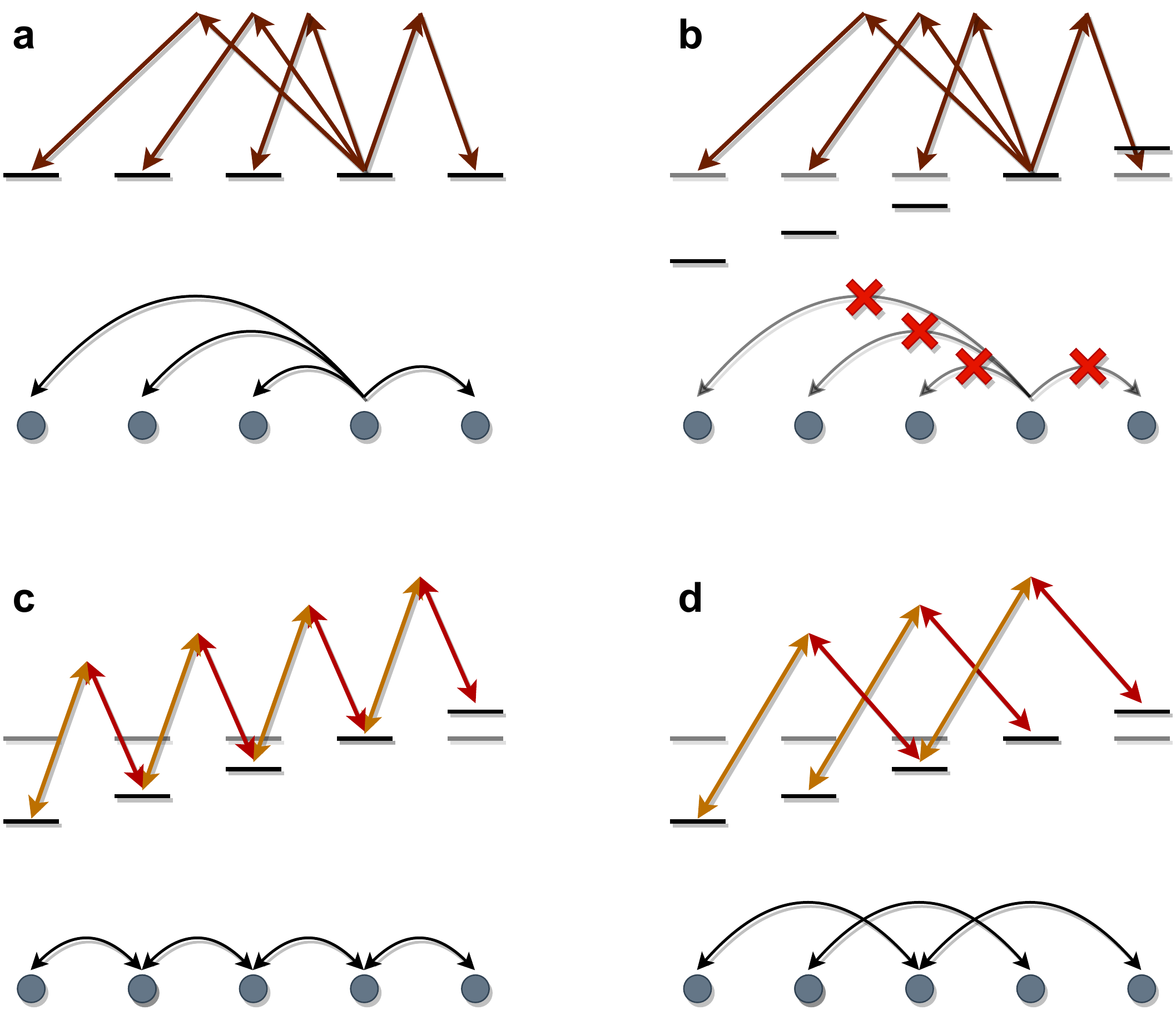}
	\caption{Coupling selectivity using an external gradient. Here black lines represent equal-excitation energy levels of the ions, while arrows represent the two-photon hopping interactions between ions. (a) In the absence of an external gradient, the hopping interaction is resonant for any ion pair. (b) By imposing an external gradient, all hopping terms are moved off resonance and therefore suppressed. (c-d) Coupling is then selectively reinstated by adding another driving frequency that bridges the gap induced by the gradient. Here either a nearest-neighbor (c) or next-nearest neighbor (d) coupling is produced through choice of driving frequencies.
	}
	\label{experiment}
\end{figure}

To do so, an external (e.g. magnetic) field  gradient is added along the ion-chain, such that the transition frequency between adjacent ions differs by $\Delta$. In order to selectively couple ions which are $n$ sites apart we drive the ions with four frequencies, composed of the frequency pairs $\omega_{b,\pm}=\omega_0+\left(\nu+\xi_b\right)\pm\frac{\Delta n}{2}$ and $\omega_{r,\pm}=\omega_0-\left(\nu+\xi_r\right)\pm\frac{\Delta n}{2}$. The pair $\omega_{b,\pm}$ couples an $n$-site hop resonantly, mediated by the blue sideband. Similarly, the pair $\omega_{r,\pm}$ couples the same hop, mediated by the red sideband. As in the MS interaction, both sidebands are employed in order to mitigate temperature-dependent effects. In order to keep the pair creation/annihilation term non-resonant we use $\xi_b=\xi+\epsilon$ and $\xi_r=\xi-\epsilon$.

Figure \ref{experiment} illustrates the transition from all-to-all coupling in the absence of a gradient field (a) to a complete suppression of coupling due to the gradient (b) and the selective resurrection of coupling by introducing the resonant sideband modulation (c,d).

\section{Derivation}
We outline the derivation of Eq. \eqref{genform}. We focus only on a single $H_n$ term, and later comment on the generalization to a summation of these terms. In the absence of an external driving field, the Hamiltonian of $N$ trapped ions in a magnetic field gradient is given by:
\begin{equation}\label{H0}
    H_{0}=\frac{\hbar}{2}\sum_{k=1}^{N}\left(\omega_{0}+k\Delta\right)\hat{\sigma}_{k}^{z}+\sum_{l=1}^{N}\hbar\nu_{l}\left(\hat{a_l}^{\dagger}\hat{a_l}+\frac{1}{2}\right).
\end{equation}
Here $\hbar\Delta$ is the transition energy difference between adjacent ions due to the field gradient, and $\nu_l$ is the frequency of the $l$-th normal-mode with the annihilation operator $a_l$. In our derivations below we will assume coupling to a single motional mode, the COM mode (with a frequency $\nu=\nu_1$ and a Lamb-Dicke parameter $\eta=\eta_1$), which couples equally to all ions in the ion-chain. This assumption can be relaxed \cite{shapira2019theory}, as will be shown in later discussions. 

For each $n$ we apply a four-tone field composed of the frequency pairs $\omega_{b,\pm}=\omega_0+\left(\nu+\xi_b\right)\pm\frac{\Delta n}{2}$ and $\omega_{r,\pm}=\omega_0-\left(\nu+\xi_r\right)\pm\frac{\Delta n}{2}$, Rabi frequencies $\Omega_b$ and $\Omega_r$, and phases $\phi_{b,\pm}=\pm\phi/2$ and $\phi_{r,\pm}=\pm\left(\phi+\pi\right)/2$. In addition $\xi_b$ and $\xi_r$ are chosen such that pair creation/annihilation transitions between all ion-pairs are detuned from resonance. 

Specifically, $\epsilon=\left(\xi_b-\xi_r\right)/2$ is chosen such that the detuning from any pair creation/annihilation resonances, i.e the transition frequency of any two states which differ by two excitations, is large compared to the effective coupling $\eta^2\Omega_r\Omega_b/\xi$ \cite{sorensen2000entanglement,shaniv2018toward}. Thus, the red and blue sideband pairs contribute to the evolution independently.

We first focus on the interaction mediated by the blue sideband, which is due to the pair $\omega_{b,\pm}$. In a frame rotating with respect to $H_0$ above, this interaction is described by, 
\begin{equation} \label{VI}
    V_I = i\hbar\eta\Omega_{b} a^{\dag}\cos\left(\frac{n\Delta}{2}t+\frac{\phi}{2}\right)\sum_{k=1}^N\sigma_{k}^{+}e^{i\left(k\Delta-\xi_b\right)t}+h.c,
\end{equation}
This expression is valid in terms of a rotating wave approximation by assuming that $\Omega_b/\nu$ and $\Omega_b/\omega_0$ are small, and in leading order in $\eta$.

The Hamiltonian in Eq. \eqref{VI} cannot be solved analytically. However, its resulting evolution operator, $U\left(t\right)$, can be approximated by using the leading terms in a Magnus expansion \cite{magnus1954onthe,blanes2010pedagogical}, $U=\exp\left[-\frac{i}{\hbar}\chi\left(t\right)\right]=\exp\left[-\frac{i}{\hbar}\sum_n \chi_n\left(t\right)\right]$. Since $\chi_n\propto\left(\frac{\eta\Omega_b}{\xi}\right)^n$, we are satisfied with terminating the expansion at the second order, which (as we show below) provides the leading order resonant terms.

The evolution due to $\chi$ is stroboscopic with a fundamental period $T$, i.e that $\chi\left(kT\right)= k T H_\text{eff}$, with $k\in\mathbb{Z}$ and $H_\text{eff}$ an effective time-independent Hamiltonian. In the limit $T\rightarrow 0$ the derived Hamiltonian approaches the target Hamiltonian at all times.

In first order the expansion reads,
\begin{equation}\label{Om1}
    \chi_1=\int_0^T dt V_I\left(t\right).
\end{equation}
By choosing $T\Delta =4\pi m$ and $T\xi_b = 2\pi M_b$, with $m,M_b\in\mathbb{Z}$, we arrive at $\chi_1\left(T\right)=0$, trivially satisfying the stroboscopic condition.

In the next order we decompose $\chi_2$ to a hopping term and a  rotation around the $z$-axis. The expansion is then given as,
\begin{equation}\label{Om2}
\begin{split}
    \chi_2&=-\frac{i}{\hbar}\int_0^T dt_1 \int_0^{t_1} dt_2 \left[V_I\left(t_1\right),V_I\left(t_2\right)\right]=\chi_{2,\text{h}}+\chi_{2,z}\\
    \chi_{2,\text{h}}&=iT\hbar^{2}\eta^{2}\Omega_{b}^{2}\sum_{k}\frac{1}{2\xi_b-2\Delta\left(k+\frac{n}{2}\right)}\sigma_{k+n}^{+}\sigma_{k}^{-}e^{i\phi}+h.c\\
    \chi_{2,z}&=iT\hbar^{2}\eta^{2}\Omega_{b}^2\sum_{k}\frac{\xi_b-k\Delta}{\left(\xi_b-k\Delta\right)^{2}-\left(\frac{n\Delta}{2}\right)^{2}}\sigma_{k}^{z}\left(a^{\dag}a+\frac{1}{2}\right).
\end{split}
\end{equation}
which can be translated to an effective Hamiltonian,
\begin{equation} \label{Hterms}
    H_{\text{eff}}=H_{h}+H_{z}+H_{h}^{\nabla}+H_{z}^{\nabla},
\end{equation}
with corrections that scale as $\left(\frac{N\Delta}{\xi_b}\right)^2$. The first two terms of this Hamiltonian are homogeneous. The first term represents the desired hopping interaction (hence the subscript $h$), while the second is effectively equivalent to a temperature-dependent global magnetic field in the $z$-direction (hence the subscript $z$),
\begin{equation}\label{Hterms_homo}
    \begin{split}
    H_{h}&=\hbar\Omega_{n,b}\sum_{k}\sigma_{k+n}^{+}\sigma_{k}^{-}e^{i\phi}+h.c\\
   H_{z}&=2\hbar\Omega_{n,b}\left(a^{\dag}a+\frac{1}{2}\right)\sum_{k}\sigma_{k}^{z},
    \end{split}
\end{equation}
with $\Omega_{n,b}=\frac{\eta^{2}\Omega_{0}^{2}}{2\xi_b}$. 

Since $H_h$ is excitation preserving, by initializing the system to an eigenstate of $\sum_k \sigma_{k}^{z}$, i.e. to a state with a well defined number of excitations, $H_z$ is reduced to a global phase and can be ignored. For these initial states a two-tone driving field suffices.

Both $H_h$ and $H_z$ are proportional to $\Omega_{n,b}$, however the former also depends on the phase $\phi$. This enables the use of the red sideband pair, $\omega_{r,\pm}$, in order to eliminate $H_{z}$ entirely while maintaining $H_h$. To this end, we choose $\xi_r$ and $\Omega_r$ such that $\Omega_{n,r}=-\Omega_{n,b}$, and $\phi_{r,\pm}=\pm\frac{\phi+\pi}{2}$. Thus the combination of the two pairs yields $H_{h}\rightarrow2H_{h}$ and $H_{z}\rightarrow0$.

The two latter terms in Eq. \eqref{Hterms}, corresponding to the two terms in Eq. \eqref{Hterms_homo}, are
\begin{equation} \label{Hterms_del}
    \begin{split}
    H_{h}^{\nabla}&=\hbar\Omega_{n}\frac{\Delta}{\xi_b}\sum_{k}\left(k+\frac{n}{2}\right)\sigma_{k+n}^{+}\sigma_{k}^{-}e^{i\phi}+h.c\\
    \\H_{z}^{\nabla}&=2\hbar\Omega_{n}\left(a^{\dag}a+\frac{1}{2}\right)\frac{\Delta}{\xi_b}\sum_{k}k\sigma_{k}^{z}.
    \end{split}
\end{equation}
These terms are gradient inhomogeneous terms, i.e they are not tranlationally invariant, and vanish in the limit $\Delta\rightarrow0$. By setting $\xi_b\gg N\Delta$, $H_h^\nabla$ and $H_z^\nabla$ become negligible and we obtain a homogeneous effective Hamiltonian. Furthermore, by using the red sideband pair as described above $H_{h}^{\nabla}$ is eliminated entirely as well. 

The remaining $H_z^\nabla$ is more difficult to eliminate in the non-adiabatic regime. If the ion chain is cooled to the ground state, its contribution to the effective Hamiltonian is simplified to $H_{z}^{\nabla} \rightarrow \hbar\frac{\eta^2\Omega_0^2}{2\xi_b^2}\Delta\sum_k k\sigma^z_k$, which has the same form as that of the external gradient field in \eqref{H0}. Hence, by making a small correction to the addressing field frequencies the ground-state contribution of $H_z^\nabla$ is eliminated.

Thus, at the appropriate limits, the four-tone frequency drive yields the effective Hamiltonian,
\begin{equation}\label{Heff}
    H_\text{eff}=2\hbar\Omega_n\sum_{k=1}^N\sigma_{k+n}^{+} \sigma_{k}^{-}e^{i\phi}+h.c.
\end{equation}
The interaction may be made time-dependent by detuning the two-photon transition, $\omega_{b,\pm}\rightarrow\omega_{b,\pm}\pm\delta/2$, with $\delta\ll\Delta$ (similarly for $\omega_{r,\pm}$). We obtain an off-resonant coupling, which manifests in a time variation of the hopping phase: $\sigma_{k+n}^{+}\sigma_{k}^{-}e^{i\phi}\rightarrow\sigma_{k+n}^{+}\sigma_{k}^{-}e^{i\left(\phi+\delta t\right)}$. With this transformation we obtain the general form of $H_n$ from Eq. \eqref{genform}. 

We further note that more generally the Peierls phase can be changed in time in whatever way one wishes by changing the appropriate driving field phases, as long as all spectral component of the phase dynamics are much smaller than $\Delta$. Constant detuning, implying a linear change in time of the phase, is a specific instance of such phase dynamics. As an example, by periodically modulating the phase, the presence of AC magnetic fluxes can be realized.

For any additional hopping term, $H_{n^\prime}$, another set of drive parameters, $\{\Omega_{0}^\prime,\xi^\prime,\phi^\prime,\delta^\prime\}$ is added, but with the further requirement that for any $m=0,1,...,N-1$ our choices satisfy $||\xi^\prime-\xi|-m\Delta|\gg\texttt{max}\{\Omega_n,\Omega_n^\prime\}$, in order to avoid any unintended cross-term resonances. This requirement should hold independently for any two terms, and can most simply be fulfilled when $|\xi-\xi^\prime|>|N\Delta|$.


\section{Some target models}
In this section we briefly explore two models that can be simulated with our method and suggest several additional models. In order to clearly display the salient features of the technique, we first focus on the analytically solvable 1D Aharonov-Bohm ring. We then discuss a triangular spin ladder model exhibiting geometric frustration. Despite the simplicity of this model, it gives rise to a rich phase diagram and interesting physical phenomena. 


\subsection{The discrete 1D Aharonov-Bohm ring}
For $N$ ions, turning on the $n=1$ and $n=N-1$ interaction terms in Eq. \eqref{genform} generates a 1D ring lattice. Adding phase terms $\phi_1=-\phi_{N-1}=2\pi\Phi/N$ simulates a magnetic flux $\Phi$ penetrating the ring, as is shown in Fig. \ref{mainresult}, with the Hamiltonian $H_r$ expressed in Eq. \eqref{ring}. As the Hamiltonian is excitation preserving, inside an excitation eigenspace $H_r$ can be mapped directly via the Jordan-Wigner transformation \cite{shankar2017exact} to fermions on a ring lattice threaded by a magnetic field:
\begin{equation}\label{fermiring}
    H_f = \sum_{i=1}^N e^{i2\pi\Phi/N}\psi^\dagger_i\psi_{i+1} + h.c.,
\end{equation}
with periodic (antiperiodic) boundary conditions for an odd (even) number of excitations. $H_f$ is the spinless fermion 1D tight-binding model \cite{viefers2004quantum}. 

The magnetic flux threading the ring may give rise to a persistent current due to the Aharonov-Bohm effect, which survives even in the presence of impurities in the chain \cite{butiker1983josephson,hofai1988persistent}. 
In order to observe this effect, one can prepare an initial state with a position occupation distribution that will rotate around the ring without diffusing. The singly-excited subspace of $H_f$ is spanned by plane-wave eigenstates $\ket{k}=\frac{1}{\sqrt{N}}\sum_n e^{ink/N}\psi^\dagger_n\ket{0}$ with energies $E_k=2\cos\left(\frac{2\pi}{N}\left(k+\Phi\right)\right)$.
While each of these waves uniformly occupies all ions along the chain, we can initialize the system in a wave packet state with a position-dependent occupation: $\ket{\psi_\text{W.P}\left(t=0\right)}=\left(\ket{k}+e^{i\varphi}\ket{k-1}\right)/\sqrt{2}$ \cite{antonio2013transport}. In this state the probability to occupy the $n$-th site is
\begin{equation}
    \left|\braket{n}{\psi_\text{W.P}\left(0\right)}\right|^2\propto 1+\cos\left(\frac{2\pi n}{N}-\varphi\right).\label{eqPacketProb}
\end{equation}
Equation \eqref{eqPacketProb} shows that $\ket{\psi_\text{W.P}}$ is a wave-packet, with $\varphi$ determining its position on the ring. The wave-packet's evolution can be described by the evolution of $\varphi$. The state will evolve according to  $\varphi\left(t\right)=\varphi\left(0\right)+v\left(\Phi\right)t$, with
\begin{equation}
    v\left(\Phi\right)=-4\Omega\sin\left(\frac{\pi}{N}\right)\sin\left(\frac{2\pi}{N}\left(\Phi+k-\frac{1}{2}\right)\right).\label{vPhi}    
\end{equation}
That is, the packet rotates around the ring at a constant, flux-dependent, angular velocity. As expected, at the large $N$ limit, we obtain $v\left(\Phi\right)\propto\Phi+k$. 

\begin{figure}
    \includegraphics[width=\columnwidth]{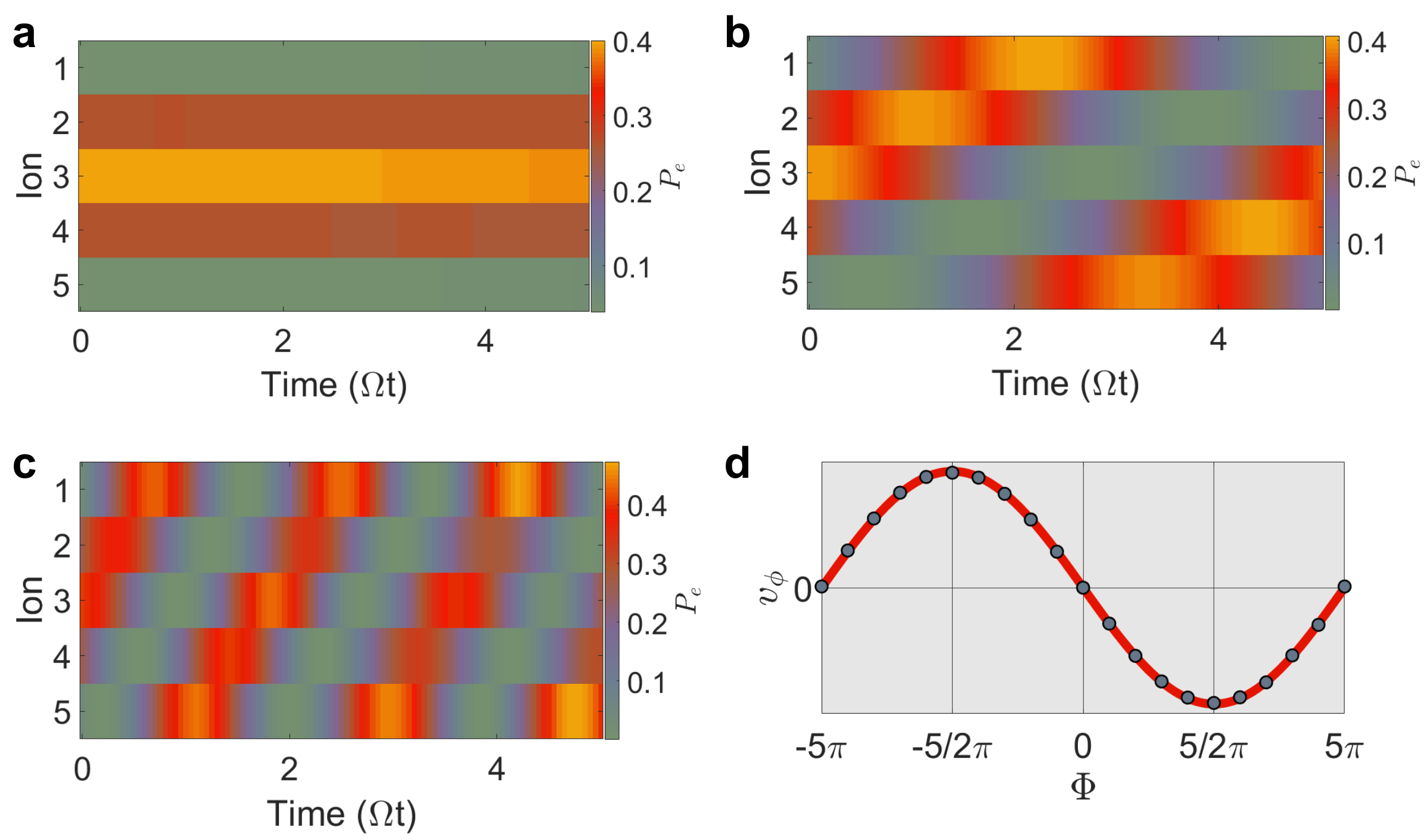}
	\caption{Time evolution in the Aharonov-Bohm ring. Turning on the $n=1$ and $n=N-1$ interaction terms in Eq. \eqref{genform} generates a closed ring threaded by a magnetic flux $\Phi=(N-1)
	\phi_1 - \phi_{N-1}$. This simulation of the time evolution of a 5 ion chain for fluxes (a) $\Phi=0$, (b) $\Phi = \pi/2$, and (c) $\Phi = 2\pi$, shows the flux-dependent propagation dynamics of a wave-packet. Local spin excitation probability $P_e=\frac{1}{2}\left( 1+\left\langle\sigma^z\right\rangle \right)$ is color coded. The excitation, which is static in the absence of flux (a), rotates around the ring at a constant velocity when flux is added. This rotation is a manifestation of persistent current in the magnetically driven ring. The ion dynamics are integrated directly from Eq. \eqref{VI} with the appropriate choice of driving parameters as detailed in this manuscript. (d) Comparison of simulation results for ion dynamics (blue) and Eq. \eqref{vPhi} (red) for the flux-dependent angular velocity of an excitation.
	}
	\label{absim}
\end{figure}

Figure \ref{absim}(a)-(c) shows simulations of the evolution of $\ket{\psi_\text{w.p}}$ with $k=0$, $N=5$ and different values of $\Phi$, integrated from Eq. \eqref{VI} with the appropriate driving field parameters. Indeed the wave packet circles around the ring with a flux-dependent velocity, exhibiting a persistent current. Figure \ref{absim}(d) compares the observed angular velocity of the wave packet with Eq. \eqref{vPhi}, showing an excellent agreement.

The Aharonov Bohm ring can be used to observe Bloch oscillations. For particles in a 1D periodic structure, the addition of a constant uniform force generates an oscillatory motion rather than unidirectional acceleration \cite{bloch1929quantenmechanik}. In a 1D ring such a force can be created by threading the ring with a time-dependent magnetic flux \cite{butiker1983josephson}. An excitation, rather than encircling the ring with a constant acceleration, will oscillate locally. The effect can be naturally incorporated using our technique, taking advantage of the ability to generate a \emph{time-varying} synthetic gauge field using off-resonant driving pairs within the coupling scheme of the AB ring, as previously described.

\subsection{Triangular Spin Ladder}

The Aharonov-Bohm ring can be mapped onto a free fermion model via the Jordan-Wigner transformation, and is thus easily solvable. However, spin interactions beyond nearest-neighbor can only be mapped onto interacting fermion models, and accordingly generate complex dynamics and phases which are often challenging for classical computation techniques. As an example, we briefly discuss the triangular ladder, a simple Hamiltonian that can be easily implemented using our technique, but which nonetheless manifests complex behavior. 

\begin{figure}
    \includegraphics[width=\columnwidth]{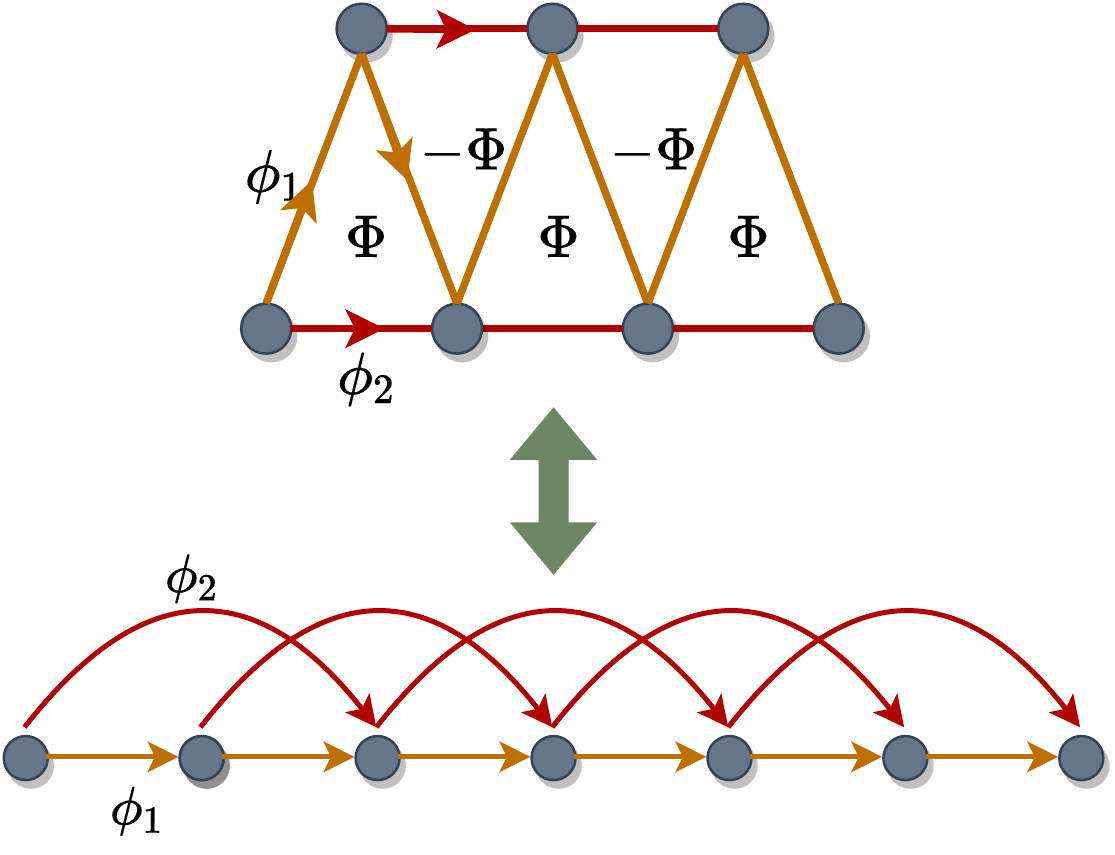}
	\caption{The triangular spin ladder. By turning on both nn and nnn couplings, an effective triangular or zigzag ladder coupling geometry is generated. The triangular layout can easily lead to frustration, which can be understood as a competition of the order imposed by the different coupling terms. Corresponding coupling phases $\phi_1$ and $\phi_2$ create a synthetic gauge field representing a magnetic flux that alternates in sign between plaquettes, with a gauge-invariant phase $\Phi=\phi_2-2\phi_1$. 
	}
	\label{triangle}
\end{figure}

Activation of the $n=1$ and $n=2$ terms in Eq. (\ref{genform}) generates a nearest neighbor (nn) and next-nearest neighbor (nnn) interaction spin Hamiltonian:
\begin{equation}
    H_{tl}=\sum_i \sigma^+_i\left(J_1e^{i\phi_1}\sigma^-_{i+1} +
    J_2 e^{i\phi_2} \sigma^-_{i+2}\right) + h.c.
\end{equation}
where boundary conditions are open. Such Hamiltonians can be graphically represented by triangular ladders in which rungs and rails represent nn and nnn interactions accordingly, as pictured in Figure \ref{triangle}. Due to the competition between nn and nnn terms, geometrically viewed as the competition of interactions inside each triangle, models of this sort are frustrated and thus give rise to a relatively rich phase diagram \cite{amico2008entanglement,diep2013frustrated,majumdar1969next}. $H_{tl}$ is gauge invariant under the transformation $\phi_1\rightarrow \phi_1+\varphi$, $\phi_2\rightarrow \phi_2+2\varphi$ for any $\varphi$; this is equal to the gauge transformation $\sigma_k^+\rightarrow e^{ik\varphi}\sigma_k^+$.   

For trivial interaction phases $\phi_2=2\phi_1$ and antiferromagnetic nnn interactions $(J_2>0)$, $H_{tl}$ represents the one-dimensional frustrated $XY$ chain model for spin-$\frac{1}{2}$. This model is a paradigmatic example of frustration \cite{diep2013frustrated}. It supports a variety of phases, notably including an exotic chiral-ordered phase \cite{furukawa2012ground,furukawa2010chiral,hikihara2008vector,nersesyan1998incommensurate}. The system phase depends on $j=J_2/J_1$, which can be fully controlled using the techniques outlined in this manuscript. 
At $j=\frac{1}{2}$, also known as the Majumdar-Ghosh point, the ground state is an exactly solvable dimerized state \cite{majumdar1969next}. The control and flexibility of trapped ion systems may enable generation of these unique phases and direct measurement of their order parameters \cite{islam2011onset}, their entanglement properties \cite{brydges2019} and their excitation dynamics \cite{Jurcevic2014}. 

By choosing nontrivial values for $\phi_1,\phi_2$ an additional synthetic gauge field representing a  \emph{staggered flux} is added to the Hamiltonian. Each triangle is pierced by a gauge-invariant magnetic flux $\pm\Phi=\phi_2-2\phi_1$, with the flux alternating signs between neighbouring plaquettes. Similar triangular ladder models with synthetic flux fields have been suggested and implemented in neutral atom systems \cite{anisimovas2016semisynthetic,suszalski2016different,an2018engineering,cabedo2020effective}. Figure \ref{triangle} illustrates the connectivity and staggered flux for this model.

\subsection{Rectangular lattice}

In the 2D examples discussed in this manuscript so far, lattices were either triangular or helical, and not rectangular. This is due to the fact that a strictly 2D rectangular lattice cannot be reduced to the form given by Eq. (\ref{genform}); placing qubits on the lattice, the nearest-neighbor coupling scheme would create a link between the last qubit in row $k$ and the first qubit in row $k+1$, violating the lattice geometry. 

\begin{figure}
    \includegraphics[width=\columnwidth]{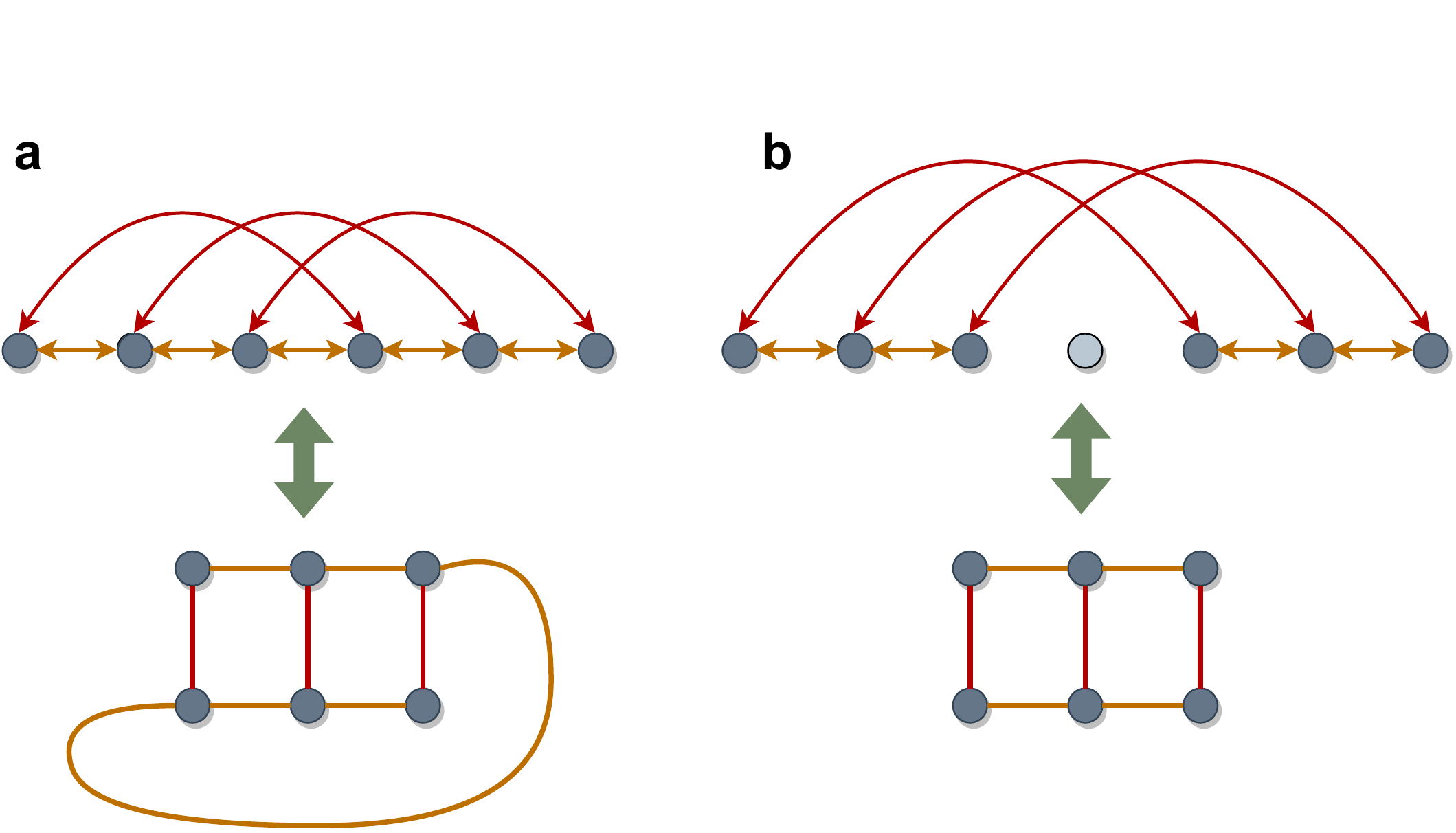}
	\caption{Spacers and rectangular ladders. (a) Placing the spins on a rectangular chain and activating the $n=1$ and $n=m$ terms generates a rectangular lattice with additional unwanted cross-row terms. (b) By adding spacer ions, initialized in a state which is uncoupled to the spin dynamics, and by replacing the $n=m$ term with a $n=m+1$ term, the unwanted coupling is corrected, giving an exact rectangular coupling geometry.  
	}
	\label{rect}
\end{figure}

This can be remedied by interrupting the nearest-neighbor interaction chains, represented by the rows of the lattice, through insertion of an auxiliary passive ion. The auxiliary ion acts as a spacer, generating an effective jump in the gradient for the active ions. This ion can be of a different isotope or species, but is most easily chosen to be an identical ion that is either strongly light shifted by an individual addressing beam or prepared in any state outside the qubit subspace. 

In the simplest example, a single spacer ion in the middle of a chain of $2N+1$ ions, along with interaction terms $H_1$ and $H_{N+1}$, would generate the rectangular spin ladder, shown in Figure \ref{rect}. With the addition of more spacer ions, more rows could be added to this array, effectively creating a complete rectangular lattice. 

This lattice can then be curled into a cylinder with the activation of an additional term. For $N=h\cdot (w+1)$ ions, where $h,w\in\mathbb{Z}$, activating the nn, $w+1$ and $N-1-(w+1)$ terms generates an $h\times w$  rectangular lattice on a cylinder. The curled dimension can be threaded by a flux, determined by phases of the non-nn terms.

\subsection{Additional geometries}
There are a number of other geometries which can be generated in a straightforward manner using our technique. We briefly mention several more examples: the M{\"o}bius ring, the cylindrical helix, and the torus, all illustrated in Fig. \ref{models}.

Turning on the terms $H_1$ and $H_{N/2}$ generates a $2\times N/2$ rectangular lattice with an additional connection of site $n-1$ to site $n$, as is shown in Fig. \ref{rect}a for $n=3$. Activating in addition the term $H_{N-1}$ forms a M{\"o}bius ring, shown in Fig. \ref{models}d. Such a system may be used to study topological effects in non-trivial geometries \cite{zhao2009mobius1,guo2009mobius2,ningyuan2015mobius3,yakubo2003}.

For $N=w\cdot h$ where $w,h\in\mathbb{Z}$, turning on the terms $H_1$ and $H_w$ in Eq. \eqref{genform} generates the cylindrical helix in Fig. \ref{models}e, with $w$ sites per loop and height $h$. Furthermore, adding terms $\Omega_{N-1}$ and $\Omega_{N-w}$ induces periodic boundary conditions, resulting in the torus seen in Fig \ref{models}f. Setting the phases $\phi_1=-\phi_{N-1}$ and $\phi_w=-\phi_{N-w}$ gives rise to two independent fluxes penetrating the torus, $\Phi_1=L\phi_w$ (green arrow) and $\Phi_2=W\phi_1-\phi_w$ (red arrow). Such a system may be used for the study of the quantum Hall effect in the thin torus limit \cite{bergholtz2005half,bernevig2012thintorus}.

In Appendix A we show a simple resource efficient implementation of our method, that can be used for the realization of some of the models above.

The ideas presented here may be taken even further by adopting the powerful neutral atoms quantum simulation concept of synthetic dimensions \cite{Boada2012,celi2016,mancini2015observation}. In neutral atom quantum simulators, extraneous internal degrees of freedom of the atom are used to represent additional lattice sites. For instance, a 1D system of atoms can be used to represent a 2D lattice, where the supplementary dimension is embodied by additional internal states of each atom. In such a case, engineered spin-orbit coupling can be used to drive a synthetic gauge field. In trapped ions, the additional Zeeman or hyperfine states may present a similar possibility, further extending ion chain simulations to an additional dimension.

\subsection{Spatially varying potentials}

In ion chains, it is possible to generate site-dependent energy shifts, $H_v=\frac{1}{2}\hbar\sum_k V_k\sigma^z_k$, by using individual addressing beams, nonuniform global beams, magnetic fields or other spatially varying fields. Under the condition  $V_k\ll\Delta$, $H_v$ can be applied in parallel to our technique. As $H_v$ commutes with $H_0$, it can be simply added to the interaction picture Hamiltonian, $V_I\rightarrow V_I+H_v$. In this instance, $H_v$ can be interpreted as a spatially varying potential on the lattice. By choosing random site-dependent shifts, disorder is added to the system. Disorder can give rise to localization \cite{anderson1958localization}, which can thus be studied in the variety of contexts presented in the paper. 

\section{Scaling up}

While implementation of the technique we propose in small scale simulators should be straightforward, approaching larger, quantum-advantage (NISQ) sized simulators \cite{preskill2018quantum} seems feasible yet more demanding. Here we discuss possible challenges in the implementation of the technique in large quantum simulators. 

The technique, as presented, calls for constant differences in resonance frequencies between neighboring ions. This can be exactly achieved with a spatially uniform gradient only to the extent that the ions are spaced equidistantly. However, many linear ion traps currently use a harmonic axial potential, in which ions are not equidistant \cite{james2000quantum}. Higher order components of the external field can be engineered to meet this issue, albeit with increasing experimental complexity. Nevertheless, several groups have constructed - or are in the process of constructing - anharmonic traps, with the stated purpose of trapping ions equidistantly \cite{pagano2018cryogenic,davoudi2019towards,lin2009,doret2012}. Anharmonic traps are likely to become more useful in future ion trap simulators and computers, due to their advantage in maintaining high inter-ion spacing, critical for preventing cross-talk in addressing and detection; their resistance to transitions from linear to zig-zag crystal configurations; and their suppression of inhomogeneous quadrupole shifts. Our proposal is best suited for such traps. 

Another hurdle for large scale implementation could be the inverse relation between coupling strength and number of ions, keeping driving field intensity constant. We define adiabaticity parameters $\alpha=\xi/\left(\Delta N\right)$ and $\beta=\Delta/\frac{\eta^2\Omega_0^2}{2\xi}$. In the low $\alpha$ limit the inhomogeneous contribution of the gradient field to the Hamiltonian becomes prominent (although strongly suppressed by the double sideband frequency configuration), and for $\alpha<1/2$ some ions in the chain may even be resonantly driven on the sideband transition. Similarly, for small values of $\beta$, non-resonant Hamiltonian terms, which are ideally completely suppressed, can play a significant role in the dynamics. Keeping these adiabaticity parameters constant, the effective coupling strength is $\frac{\eta^2\Omega_0^2}{2\xi}=\frac{\eta\Omega_0}{\sqrt{2\alpha\beta N}}$. Since we assume coupling only to the COM mode, the Lamb-Dicke parameter also decreases with ion number: $\eta=\frac{\eta_1}{\sqrt{N}}$, where $\eta_1$ is the single-ion Lamb-Dicke parameter. Hence, overall, the coupling strength decreases as $\frac{1}{N}$, as opposed to the normal COM MS degradation of $\frac{1}{\sqrt{N}}$. In essence, the additional penalty emerges from the requirement to preserve spectral spacing in presence of the gradient field. The degradation can be remedied by increasing the field intensity or by coupling to many modes rather than just the COM, which is not only convenient but necessary when using radial modes in large simulators.

The radial modes of linear ion traps bunch up when adding more ions, and consequently the radial COM mode cannot be spectrally resolved. Using these modes as interaction mediators will thus inevitably generate coupling to a multitude of modes. Even so, the use of radial modes can be advantageous and is compatible with our proposal, up to a modification of the effective coupling, in analogy with \cite{porras2004effective} (see Appendix B for more details). 

In contrast, the axial COM mode remains spectrally separated from all other modes independently of the number of ions, and thus is in this sense ideal for the proposed technique. Nevertheless, axial modes come with their own disadvantages. Working with large ion chain requires very low axial trap frequencies in order to avoid the crystal zig-zag transition, or buckling, in the middle of the chain. Low axial frequencies lead to higher carrier coupling, temperatures, and heating rates, limiting simulation fidelity. These problems can be strongly mitigated by using anharmonic trapping potentials, which are beneficial for our proposal as stated above.

\section{Summary}
In conclusion, we have introduced and explored a technique that realizes a variety of spin Hamiltonians in trapped ion chains. The range of implementable Hamiltonians includes spin lattices with dimension larger than one, closed boundary conditions, rectangular and triangular lattices, and full control of nearest and next-nearest neighbor couplings. Furthermore, the technique provides a means to realize static and time-varying synthetic gauge fields in ion chains. This is done using a global driving field of uniform intensity, with no need for individual addressing or for dynamic control, and a static external field gradient along the ion chain. 
The advanced tools that have already been developed for trapped ion chains, including preparation of highly entangled states  \cite{friis2018observation} or measurement of observables of interest such as entanglement entropy \cite{brydges2019}, encourages us to believe that ion chain analog simulation can break new ground in territories which have been considered outside its purview, such as simulation of 2D topological phenomena. The tools we present here are a step towards this direction.

\paragraph{Acknowledgements---}
\begin{acknowledgments}
We thank Ori Alberton for helpful discussions. This work was supported by the Israeli Science Foundation, the Israeli Ministry of Science Technology and Space and the Minerva Stiftung.
\end{acknowledgments}

\bibliographystyle{apsrev4-1}
\bibliography{jul4sub}

\begin{thebibliography}{98}%
\makeatletter
\providecommand \@ifxundefined [1]{%
 \@ifx{#1\undefined}
}%
\providecommand \@ifnum [1]{%
 \ifnum #1\expandafter \@firstoftwo
 \else \expandafter \@secondoftwo
 \fi
}%
\providecommand \@ifx [1]{%
 \ifx #1\expandafter \@firstoftwo
 \else \expandafter \@secondoftwo
 \fi
}%
\providecommand \natexlab [1]{#1}%
\providecommand \enquote  [1]{``#1''}%
\providecommand \bibnamefont  [1]{#1}%
\providecommand \bibfnamefont [1]{#1}%
\providecommand \citenamefont [1]{#1}%
\providecommand \href@noop [0]{\@secondoftwo}%
\providecommand \href [0]{\begingroup \@sanitize@url \@href}%
\providecommand \@href[1]{\@@startlink{#1}\@@href}%
\providecommand \@@href[1]{\endgroup#1\@@endlink}%
\providecommand \@sanitize@url [0]{\catcode `\\12\catcode `\$12\catcode
  `\&12\catcode `\#12\catcode `\^12\catcode `\_12\catcode `\%12\relax}%
\providecommand \@@startlink[1]{}%
\providecommand \@@endlink[0]{}%
\providecommand \url  [0]{\begingroup\@sanitize@url \@url }%
\providecommand \@url [1]{\endgroup\@href {#1}{\urlprefix }}%
\providecommand \urlprefix  [0]{URL }%
\providecommand \Eprint [0]{\href }%
\providecommand \doibase [0]{http://dx.doi.org/}%
\providecommand \selectlanguage [0]{\@gobble}%
\providecommand \bibinfo  [0]{\@secondoftwo}%
\providecommand \bibfield  [0]{\@secondoftwo}%
\providecommand \translation [1]{[#1]}%
\providecommand \BibitemOpen [0]{}%
\providecommand \bibitemStop [0]{}%
\providecommand \bibitemNoStop [0]{.\EOS\space}%
\providecommand \EOS [0]{\spacefactor3000\relax}%
\providecommand \BibitemShut  [1]{\csname bibitem#1\endcsname}%
\let\auto@bib@innerbib\@empty
\bibitem [{\citenamefont {Feynman}(1982)}]{feynman1982simulating}%
  \BibitemOpen
  \bibfield  {author} {\bibinfo {author} {\bibfnamefont {R.~P.}\ \bibnamefont
  {Feynman}},\ }\href@noop {} {\bibfield  {journal} {\bibinfo  {journal}
  {International journal of theoretical physics}\ }\textbf {\bibinfo {volume}
  {21}},\ \bibinfo {pages} {467} (\bibinfo {year} {1982})}\BibitemShut
  {NoStop}%
\bibitem [{\citenamefont {Bernien}\ \emph {et~al.}(2017)\citenamefont
  {Bernien}, \citenamefont {Schwartz}, \citenamefont {Keesling}, \citenamefont
  {Levine}, \citenamefont {Omran}, \citenamefont {Pichler}, \citenamefont
  {Choi}, \citenamefont {Zibrov}, \citenamefont {Endres}, \citenamefont
  {Greiner} \emph {et~al.}}]{bernien2017probing}%
  \BibitemOpen
  \bibfield  {author} {\bibinfo {author} {\bibfnamefont {H.}~\bibnamefont
  {Bernien}}, \bibinfo {author} {\bibfnamefont {S.}~\bibnamefont {Schwartz}},
  \bibinfo {author} {\bibfnamefont {A.}~\bibnamefont {Keesling}}, \bibinfo
  {author} {\bibfnamefont {H.}~\bibnamefont {Levine}}, \bibinfo {author}
  {\bibfnamefont {A.}~\bibnamefont {Omran}}, \bibinfo {author} {\bibfnamefont
  {H.}~\bibnamefont {Pichler}}, \bibinfo {author} {\bibfnamefont
  {S.}~\bibnamefont {Choi}}, \bibinfo {author} {\bibfnamefont {A.~S.}\
  \bibnamefont {Zibrov}}, \bibinfo {author} {\bibfnamefont {M.}~\bibnamefont
  {Endres}}, \bibinfo {author} {\bibfnamefont {M.}~\bibnamefont {Greiner}},
  \emph {et~al.},\ }\href@noop {} {\bibfield  {journal} {\bibinfo  {journal}
  {Nature}\ }\textbf {\bibinfo {volume} {551}},\ \bibinfo {pages} {579}
  (\bibinfo {year} {2017})}\BibitemShut {NoStop}%
\bibitem [{\citenamefont {Zhang}\ \emph
  {et~al.}(2017{\natexlab{a}})\citenamefont {Zhang}, \citenamefont {Pagano},
  \citenamefont {Hess}, \citenamefont {Kyprianidis}, \citenamefont {Becker},
  \citenamefont {Kaplan}, \citenamefont {Gorshkov}, \citenamefont {Gong},\ and\
  \citenamefont {Monroe}}]{zhang2017observation}%
  \BibitemOpen
  \bibfield  {author} {\bibinfo {author} {\bibfnamefont {J.}~\bibnamefont
  {Zhang}}, \bibinfo {author} {\bibfnamefont {G.}~\bibnamefont {Pagano}},
  \bibinfo {author} {\bibfnamefont {P.~W.}\ \bibnamefont {Hess}}, \bibinfo
  {author} {\bibfnamefont {A.}~\bibnamefont {Kyprianidis}}, \bibinfo {author}
  {\bibfnamefont {P.}~\bibnamefont {Becker}}, \bibinfo {author} {\bibfnamefont
  {H.}~\bibnamefont {Kaplan}}, \bibinfo {author} {\bibfnamefont {A.~V.}\
  \bibnamefont {Gorshkov}}, \bibinfo {author} {\bibfnamefont {Z.-X.}\
  \bibnamefont {Gong}}, \ and\ \bibinfo {author} {\bibfnamefont
  {C.}~\bibnamefont {Monroe}},\ }\href@noop {} {\bibfield  {journal} {\bibinfo
  {journal} {Nature}\ }\textbf {\bibinfo {volume} {551}},\ \bibinfo {pages}
  {601} (\bibinfo {year} {2017}{\natexlab{a}})}\BibitemShut {NoStop}%
\bibitem [{\citenamefont {Arute}\ \emph {et~al.}(2019)\citenamefont {Arute},
  \citenamefont {Arya}, \citenamefont {Babbush}, \citenamefont {Bacon},
  \citenamefont {Bardin}, \citenamefont {Barends}, \citenamefont {Biswas},
  \citenamefont {Boixo}, \citenamefont {Brandao}, \citenamefont {Buell} \emph
  {et~al.}}]{arute2019quantum}%
  \BibitemOpen
  \bibfield  {author} {\bibinfo {author} {\bibfnamefont {F.}~\bibnamefont
  {Arute}}, \bibinfo {author} {\bibfnamefont {K.}~\bibnamefont {Arya}},
  \bibinfo {author} {\bibfnamefont {R.}~\bibnamefont {Babbush}}, \bibinfo
  {author} {\bibfnamefont {D.}~\bibnamefont {Bacon}}, \bibinfo {author}
  {\bibfnamefont {J.~C.}\ \bibnamefont {Bardin}}, \bibinfo {author}
  {\bibfnamefont {R.}~\bibnamefont {Barends}}, \bibinfo {author} {\bibfnamefont
  {R.}~\bibnamefont {Biswas}}, \bibinfo {author} {\bibfnamefont
  {S.}~\bibnamefont {Boixo}}, \bibinfo {author} {\bibfnamefont {F.~G.}\
  \bibnamefont {Brandao}}, \bibinfo {author} {\bibfnamefont {D.~A.}\
  \bibnamefont {Buell}},  \emph {et~al.},\ }\href@noop {} {\bibfield  {journal}
  {\bibinfo  {journal} {Nature}\ }\textbf {\bibinfo {volume} {574}},\ \bibinfo
  {pages} {505} (\bibinfo {year} {2019})}\BibitemShut {NoStop}%
\bibitem [{\citenamefont {Bohnet}\ \emph {et~al.}(2016)\citenamefont {Bohnet},
  \citenamefont {Sawyer}, \citenamefont {Britton}, \citenamefont {Wall},
  \citenamefont {Rey}, \citenamefont {Foss-Feig},\ and\ \citenamefont
  {Bollinger}}]{Bohnet2016}%
  \BibitemOpen
  \bibfield  {author} {\bibinfo {author} {\bibfnamefont {J.~G.}\ \bibnamefont
  {Bohnet}}, \bibinfo {author} {\bibfnamefont {B.~C.}\ \bibnamefont {Sawyer}},
  \bibinfo {author} {\bibfnamefont {J.~W.}\ \bibnamefont {Britton}}, \bibinfo
  {author} {\bibfnamefont {M.~L.}\ \bibnamefont {Wall}}, \bibinfo {author}
  {\bibfnamefont {A.~M.}\ \bibnamefont {Rey}}, \bibinfo {author} {\bibfnamefont
  {M.}~\bibnamefont {Foss-Feig}}, \ and\ \bibinfo {author} {\bibfnamefont
  {J.~J.}\ \bibnamefont {Bollinger}},\ }\href {\doibase
  10.1126/science.aad9958} {\bibfield  {journal} {\bibinfo  {journal}
  {Science}\ }\textbf {\bibinfo {volume} {352}},\ \bibinfo {pages} {1297}
  (\bibinfo {year} {2016})},\ \Eprint
  {http://arxiv.org/abs/https://science.sciencemag.org/content/352/6291/1297.full.pdf}
  {https://science.sciencemag.org/content/352/6291/1297.full.pdf} \BibitemShut
  {NoStop}%
\bibitem [{\citenamefont {Ac{\'{\i}}n}\ \emph {et~al.}(2018)\citenamefont
  {Ac{\'{\i}}n}, \citenamefont {Bloch}, \citenamefont {Buhrman}, \citenamefont
  {Calarco}, \citenamefont {Eichler}, \citenamefont {Eisert}, \citenamefont
  {Esteve}, \citenamefont {Gisin}, \citenamefont {Glaser}, \citenamefont
  {Jelezko}, \citenamefont {Kuhr}, \citenamefont {Lewenstein}, \citenamefont
  {Riedel}, \citenamefont {Schmidt}, \citenamefont {Thew}, \citenamefont
  {Wallraff}, \citenamefont {Walmsley},\ and\ \citenamefont
  {Wilhelm}}]{Acin2018}%
  \BibitemOpen
  \bibfield  {author} {\bibinfo {author} {\bibfnamefont {A.}~\bibnamefont
  {Ac{\'{\i}}n}}, \bibinfo {author} {\bibfnamefont {I.}~\bibnamefont {Bloch}},
  \bibinfo {author} {\bibfnamefont {H.}~\bibnamefont {Buhrman}}, \bibinfo
  {author} {\bibfnamefont {T.}~\bibnamefont {Calarco}}, \bibinfo {author}
  {\bibfnamefont {C.}~\bibnamefont {Eichler}}, \bibinfo {author} {\bibfnamefont
  {J.}~\bibnamefont {Eisert}}, \bibinfo {author} {\bibfnamefont
  {D.}~\bibnamefont {Esteve}}, \bibinfo {author} {\bibfnamefont
  {N.}~\bibnamefont {Gisin}}, \bibinfo {author} {\bibfnamefont {S.~J.}\
  \bibnamefont {Glaser}}, \bibinfo {author} {\bibfnamefont {F.}~\bibnamefont
  {Jelezko}}, \bibinfo {author} {\bibfnamefont {S.}~\bibnamefont {Kuhr}},
  \bibinfo {author} {\bibfnamefont {M.}~\bibnamefont {Lewenstein}}, \bibinfo
  {author} {\bibfnamefont {M.~F.}\ \bibnamefont {Riedel}}, \bibinfo {author}
  {\bibfnamefont {P.~O.}\ \bibnamefont {Schmidt}}, \bibinfo {author}
  {\bibfnamefont {R.}~\bibnamefont {Thew}}, \bibinfo {author} {\bibfnamefont
  {A.}~\bibnamefont {Wallraff}}, \bibinfo {author} {\bibfnamefont
  {I.}~\bibnamefont {Walmsley}}, \ and\ \bibinfo {author} {\bibfnamefont
  {F.~K.}\ \bibnamefont {Wilhelm}},\ }\href {\doibase 10.1088/1367-2630/aad1ea}
  {\bibfield  {journal} {\bibinfo  {journal} {New Journal of Physics}\ }\textbf
  {\bibinfo {volume} {20}},\ \bibinfo {pages} {080201} (\bibinfo {year}
  {2018})}\BibitemShut {NoStop}%
\bibitem [{\citenamefont {Bermejo-Vega}\ \emph {et~al.}(2018)\citenamefont
  {Bermejo-Vega}, \citenamefont {Hangleiter}, \citenamefont {Schwarz},
  \citenamefont {Raussendorf},\ and\ \citenamefont
  {Eisert}}]{bermejo2018architectures}%
  \BibitemOpen
  \bibfield  {author} {\bibinfo {author} {\bibfnamefont {J.}~\bibnamefont
  {Bermejo-Vega}}, \bibinfo {author} {\bibfnamefont {D.}~\bibnamefont
  {Hangleiter}}, \bibinfo {author} {\bibfnamefont {M.}~\bibnamefont {Schwarz}},
  \bibinfo {author} {\bibfnamefont {R.}~\bibnamefont {Raussendorf}}, \ and\
  \bibinfo {author} {\bibfnamefont {J.}~\bibnamefont {Eisert}},\ }\href@noop {}
  {\bibfield  {journal} {\bibinfo  {journal} {Physical Review X}\ }\textbf
  {\bibinfo {volume} {8}},\ \bibinfo {pages} {021010} (\bibinfo {year}
  {2018})}\BibitemShut {NoStop}%
\bibitem [{\citenamefont {Blatt}\ and\ \citenamefont {Roos}(2012)}]{Blatt2012}%
  \BibitemOpen
  \bibfield  {author} {\bibinfo {author} {\bibfnamefont {R.}~\bibnamefont
  {Blatt}}\ and\ \bibinfo {author} {\bibfnamefont {C.~F.}\ \bibnamefont
  {Roos}},\ }\href {\doibase 10.1038/nphys2252} {\bibfield  {journal} {\bibinfo
   {journal} {Nature Physics}\ }\textbf {\bibinfo {volume} {8}},\ \bibinfo
  {pages} {277} (\bibinfo {year} {2012})}\BibitemShut {NoStop}%
\bibitem [{\citenamefont {Monroe}\ \emph {et~al.}(2019)\citenamefont {Monroe},
  \citenamefont {Campbell}, \citenamefont {Duan}, \citenamefont {Gong},
  \citenamefont {Gorshkov}, \citenamefont {Hess}, \citenamefont {Islam},
  \citenamefont {Kim}, \citenamefont {Pagano}, \citenamefont {Richerme} \emph
  {et~al.}}]{monroe2019programmable}%
  \BibitemOpen
  \bibfield  {author} {\bibinfo {author} {\bibfnamefont {C.}~\bibnamefont
  {Monroe}}, \bibinfo {author} {\bibfnamefont {W.}~\bibnamefont {Campbell}},
  \bibinfo {author} {\bibfnamefont {L.-M.}\ \bibnamefont {Duan}}, \bibinfo
  {author} {\bibfnamefont {Z.-X.}\ \bibnamefont {Gong}}, \bibinfo {author}
  {\bibfnamefont {A.}~\bibnamefont {Gorshkov}}, \bibinfo {author}
  {\bibfnamefont {P.}~\bibnamefont {Hess}}, \bibinfo {author} {\bibfnamefont
  {R.}~\bibnamefont {Islam}}, \bibinfo {author} {\bibfnamefont
  {K.}~\bibnamefont {Kim}}, \bibinfo {author} {\bibfnamefont {G.}~\bibnamefont
  {Pagano}}, \bibinfo {author} {\bibfnamefont {P.}~\bibnamefont {Richerme}},
  \emph {et~al.},\ }\href@noop {} {\bibfield  {journal} {\bibinfo  {journal}
  {arXiv preprint arXiv:1912.07845}\ } (\bibinfo {year} {2019})}\BibitemShut
  {NoStop}%
\bibitem [{\citenamefont {Gerritsma}\ \emph {et~al.}(2010)\citenamefont
  {Gerritsma}, \citenamefont {Kirchmair}, \citenamefont {Z{\"a}hringer},
  \citenamefont {Solano}, \citenamefont {Blatt},\ and\ \citenamefont
  {Roos}}]{gerritsma2010quantum}%
  \BibitemOpen
  \bibfield  {author} {\bibinfo {author} {\bibfnamefont {R.}~\bibnamefont
  {Gerritsma}}, \bibinfo {author} {\bibfnamefont {G.}~\bibnamefont
  {Kirchmair}}, \bibinfo {author} {\bibfnamefont {F.}~\bibnamefont
  {Z{\"a}hringer}}, \bibinfo {author} {\bibfnamefont {E.}~\bibnamefont
  {Solano}}, \bibinfo {author} {\bibfnamefont {R.}~\bibnamefont {Blatt}}, \
  and\ \bibinfo {author} {\bibfnamefont {C.}~\bibnamefont {Roos}},\ }\href@noop
  {} {\bibfield  {journal} {\bibinfo  {journal} {Nature}\ }\textbf {\bibinfo
  {volume} {463}},\ \bibinfo {pages} {68} (\bibinfo {year} {2010})}\BibitemShut
  {NoStop}%
\bibitem [{\citenamefont {Kim}\ \emph {et~al.}(2010)\citenamefont {Kim},
  \citenamefont {Chang}, \citenamefont {Korenblit}, \citenamefont {Islam},
  \citenamefont {Edwards}, \citenamefont {Freericks}, \citenamefont {Lin},
  \citenamefont {Duan},\ and\ \citenamefont {Monroe}}]{Kim2010}%
  \BibitemOpen
  \bibfield  {author} {\bibinfo {author} {\bibfnamefont {K.}~\bibnamefont
  {Kim}}, \bibinfo {author} {\bibfnamefont {M.-S.}\ \bibnamefont {Chang}},
  \bibinfo {author} {\bibfnamefont {S.}~\bibnamefont {Korenblit}}, \bibinfo
  {author} {\bibfnamefont {R.}~\bibnamefont {Islam}}, \bibinfo {author}
  {\bibfnamefont {E.~E.}\ \bibnamefont {Edwards}}, \bibinfo {author}
  {\bibfnamefont {J.~K.}\ \bibnamefont {Freericks}}, \bibinfo {author}
  {\bibfnamefont {G.-D.}\ \bibnamefont {Lin}}, \bibinfo {author} {\bibfnamefont
  {L.-M.}\ \bibnamefont {Duan}}, \ and\ \bibinfo {author} {\bibfnamefont
  {C.}~\bibnamefont {Monroe}},\ }\href {\doibase 10.1038/nature09071}
  {\bibfield  {journal} {\bibinfo  {journal} {Nature}\ }\textbf {\bibinfo
  {volume} {465}},\ \bibinfo {pages} {590} (\bibinfo {year}
  {2010})}\BibitemShut {NoStop}%
\bibitem [{\citenamefont {Islam}\ \emph
  {et~al.}(2011{\natexlab{a}})\citenamefont {Islam}, \citenamefont {Edwards},
  \citenamefont {Kim}, \citenamefont {Korenblit}, \citenamefont {Noh},
  \citenamefont {Carmichael}, \citenamefont {Lin}, \citenamefont {Duan},
  \citenamefont {Wang}, \citenamefont {Freericks},\ and\ \citenamefont
  {Monroe}}]{Islam2011}%
  \BibitemOpen
  \bibfield  {author} {\bibinfo {author} {\bibfnamefont {R.}~\bibnamefont
  {Islam}}, \bibinfo {author} {\bibfnamefont {E.}~\bibnamefont {Edwards}},
  \bibinfo {author} {\bibfnamefont {K.}~\bibnamefont {Kim}}, \bibinfo {author}
  {\bibfnamefont {S.}~\bibnamefont {Korenblit}}, \bibinfo {author}
  {\bibfnamefont {C.}~\bibnamefont {Noh}}, \bibinfo {author} {\bibfnamefont
  {H.}~\bibnamefont {Carmichael}}, \bibinfo {author} {\bibfnamefont {G.-D.}\
  \bibnamefont {Lin}}, \bibinfo {author} {\bibfnamefont {L.-M.}\ \bibnamefont
  {Duan}}, \bibinfo {author} {\bibfnamefont {C.-C.~J.}\ \bibnamefont {Wang}},
  \bibinfo {author} {\bibfnamefont {J.}~\bibnamefont {Freericks}}, \ and\
  \bibinfo {author} {\bibfnamefont {C.}~\bibnamefont {Monroe}},\ }\href
  {\doibase 10.1038/ncomms1374} {\bibfield  {journal} {\bibinfo  {journal}
  {Nature Communications}\ }\textbf {\bibinfo {volume} {2}} (\bibinfo {year}
  {2011}{\natexlab{a}}),\ 10.1038/ncomms1374}\BibitemShut {NoStop}%
\bibitem [{\citenamefont {Islam}\ \emph {et~al.}(2013)\citenamefont {Islam},
  \citenamefont {Senko}, \citenamefont {Campbell}, \citenamefont {Korenblit},
  \citenamefont {Smith}, \citenamefont {Lee}, \citenamefont {Edwards},
  \citenamefont {Wang}, \citenamefont {Freericks},\ and\ \citenamefont
  {Monroe}}]{islam2013emergence}%
  \BibitemOpen
  \bibfield  {author} {\bibinfo {author} {\bibfnamefont {R.}~\bibnamefont
  {Islam}}, \bibinfo {author} {\bibfnamefont {C.}~\bibnamefont {Senko}},
  \bibinfo {author} {\bibfnamefont {W.}~\bibnamefont {Campbell}}, \bibinfo
  {author} {\bibfnamefont {S.}~\bibnamefont {Korenblit}}, \bibinfo {author}
  {\bibfnamefont {J.}~\bibnamefont {Smith}}, \bibinfo {author} {\bibfnamefont
  {A.}~\bibnamefont {Lee}}, \bibinfo {author} {\bibfnamefont {E.}~\bibnamefont
  {Edwards}}, \bibinfo {author} {\bibfnamefont {C.-C.}\ \bibnamefont {Wang}},
  \bibinfo {author} {\bibfnamefont {J.}~\bibnamefont {Freericks}}, \ and\
  \bibinfo {author} {\bibfnamefont {C.}~\bibnamefont {Monroe}},\ }\href@noop {}
  {\bibfield  {journal} {\bibinfo  {journal} {Science}\ }\textbf {\bibinfo
  {volume} {340}},\ \bibinfo {pages} {583} (\bibinfo {year}
  {2013})}\BibitemShut {NoStop}%
\bibitem [{\citenamefont {Schachenmayer}\ \emph {et~al.}(2013)\citenamefont
  {Schachenmayer}, \citenamefont {Lanyon}, \citenamefont {Roos},\ and\
  \citenamefont {Daley}}]{schachenmayer2013}%
  \BibitemOpen
  \bibfield  {author} {\bibinfo {author} {\bibfnamefont {J.}~\bibnamefont
  {Schachenmayer}}, \bibinfo {author} {\bibfnamefont {B.~P.}\ \bibnamefont
  {Lanyon}}, \bibinfo {author} {\bibfnamefont {C.~F.}\ \bibnamefont {Roos}}, \
  and\ \bibinfo {author} {\bibfnamefont {A.~J.}\ \bibnamefont {Daley}},\ }\href
  {\doibase 10.1103/PhysRevX.3.031015} {\bibfield  {journal} {\bibinfo
  {journal} {Phys. Rev. X}\ }\textbf {\bibinfo {volume} {3}},\ \bibinfo {pages}
  {031015} (\bibinfo {year} {2013})}\BibitemShut {NoStop}%
\bibitem [{\citenamefont {Richerme}\ \emph {et~al.}(2014)\citenamefont
  {Richerme}, \citenamefont {Gong}, \citenamefont {Lee}, \citenamefont {Senko},
  \citenamefont {Smith}, \citenamefont {Foss-Feig}, \citenamefont {Michalakis},
  \citenamefont {Gorshkov},\ and\ \citenamefont {Monroe}}]{Richerme2014}%
  \BibitemOpen
  \bibfield  {author} {\bibinfo {author} {\bibfnamefont {P.}~\bibnamefont
  {Richerme}}, \bibinfo {author} {\bibfnamefont {Z.-X.}\ \bibnamefont {Gong}},
  \bibinfo {author} {\bibfnamefont {A.}~\bibnamefont {Lee}}, \bibinfo {author}
  {\bibfnamefont {C.}~\bibnamefont {Senko}}, \bibinfo {author} {\bibfnamefont
  {J.}~\bibnamefont {Smith}}, \bibinfo {author} {\bibfnamefont
  {M.}~\bibnamefont {Foss-Feig}}, \bibinfo {author} {\bibfnamefont
  {S.}~\bibnamefont {Michalakis}}, \bibinfo {author} {\bibfnamefont {A.~V.}\
  \bibnamefont {Gorshkov}}, \ and\ \bibinfo {author} {\bibfnamefont
  {C.}~\bibnamefont {Monroe}},\ }\href {\doibase 10.1038/nature13450}
  {\bibfield  {journal} {\bibinfo  {journal} {Nature}\ }\textbf {\bibinfo
  {volume} {511}},\ \bibinfo {pages} {198} (\bibinfo {year}
  {2014})}\BibitemShut {NoStop}%
\bibitem [{\citenamefont {Jurcevic}\ \emph {et~al.}(2014)\citenamefont
  {Jurcevic}, \citenamefont {Lanyon}, \citenamefont {Hauke}, \citenamefont
  {Hempel}, \citenamefont {Zoller}, \citenamefont {Blatt},\ and\ \citenamefont
  {Roos}}]{Jurcevic2014}%
  \BibitemOpen
  \bibfield  {author} {\bibinfo {author} {\bibfnamefont {P.}~\bibnamefont
  {Jurcevic}}, \bibinfo {author} {\bibfnamefont {B.~P.}\ \bibnamefont
  {Lanyon}}, \bibinfo {author} {\bibfnamefont {P.}~\bibnamefont {Hauke}},
  \bibinfo {author} {\bibfnamefont {C.}~\bibnamefont {Hempel}}, \bibinfo
  {author} {\bibfnamefont {P.}~\bibnamefont {Zoller}}, \bibinfo {author}
  {\bibfnamefont {R.}~\bibnamefont {Blatt}}, \ and\ \bibinfo {author}
  {\bibfnamefont {C.~F.}\ \bibnamefont {Roos}},\ }\href {\doibase
  10.1038/nature13461} {\bibfield  {journal} {\bibinfo  {journal} {Nature}\
  }\textbf {\bibinfo {volume} {511}},\ \bibinfo {pages} {202} (\bibinfo {year}
  {2014})}\BibitemShut {NoStop}%
\bibitem [{\citenamefont {Smith}\ \emph {et~al.}(2016)\citenamefont {Smith},
  \citenamefont {Lee}, \citenamefont {Richerme}, \citenamefont {Neyenhuis},
  \citenamefont {Hess}, \citenamefont {Hauke}, \citenamefont {Heyl},
  \citenamefont {Huse},\ and\ \citenamefont {Monroe}}]{Smith2016}%
  \BibitemOpen
  \bibfield  {author} {\bibinfo {author} {\bibfnamefont {J.}~\bibnamefont
  {Smith}}, \bibinfo {author} {\bibfnamefont {A.}~\bibnamefont {Lee}}, \bibinfo
  {author} {\bibfnamefont {P.}~\bibnamefont {Richerme}}, \bibinfo {author}
  {\bibfnamefont {B.}~\bibnamefont {Neyenhuis}}, \bibinfo {author}
  {\bibfnamefont {P.~W.}\ \bibnamefont {Hess}}, \bibinfo {author}
  {\bibfnamefont {P.}~\bibnamefont {Hauke}}, \bibinfo {author} {\bibfnamefont
  {M.}~\bibnamefont {Heyl}}, \bibinfo {author} {\bibfnamefont {D.~A.}\
  \bibnamefont {Huse}}, \ and\ \bibinfo {author} {\bibfnamefont
  {C.}~\bibnamefont {Monroe}},\ }\href {\doibase 10.1038/nphys3783} {\bibfield
  {journal} {\bibinfo  {journal} {Nature Physics}\ }\textbf {\bibinfo {volume}
  {12}},\ \bibinfo {pages} {907} (\bibinfo {year} {2016})}\BibitemShut
  {NoStop}%
\bibitem [{\citenamefont {Jurcevic}\ \emph {et~al.}(2017)\citenamefont
  {Jurcevic}, \citenamefont {Shen}, \citenamefont {Hauke}, \citenamefont
  {Maier}, \citenamefont {Brydges}, \citenamefont {Hempel}, \citenamefont
  {Lanyon}, \citenamefont {Heyl}, \citenamefont {Blatt},\ and\ \citenamefont
  {Roos}}]{jurcevic2017}%
  \BibitemOpen
  \bibfield  {author} {\bibinfo {author} {\bibfnamefont {P.}~\bibnamefont
  {Jurcevic}}, \bibinfo {author} {\bibfnamefont {H.}~\bibnamefont {Shen}},
  \bibinfo {author} {\bibfnamefont {P.}~\bibnamefont {Hauke}}, \bibinfo
  {author} {\bibfnamefont {C.}~\bibnamefont {Maier}}, \bibinfo {author}
  {\bibfnamefont {T.}~\bibnamefont {Brydges}}, \bibinfo {author} {\bibfnamefont
  {C.}~\bibnamefont {Hempel}}, \bibinfo {author} {\bibfnamefont {B.~P.}\
  \bibnamefont {Lanyon}}, \bibinfo {author} {\bibfnamefont {M.}~\bibnamefont
  {Heyl}}, \bibinfo {author} {\bibfnamefont {R.}~\bibnamefont {Blatt}}, \ and\
  \bibinfo {author} {\bibfnamefont {C.~F.}\ \bibnamefont {Roos}},\ }\href
  {\doibase 10.1103/PhysRevLett.119.080501} {\bibfield  {journal} {\bibinfo
  {journal} {Phys. Rev. Lett.}\ }\textbf {\bibinfo {volume} {119}},\ \bibinfo
  {pages} {080501} (\bibinfo {year} {2017})}\BibitemShut {NoStop}%
\bibitem [{\citenamefont {Zhang}\ \emph
  {et~al.}(2017{\natexlab{b}})\citenamefont {Zhang}, \citenamefont {Hess},
  \citenamefont {Kyprianidis}, \citenamefont {Becker}, \citenamefont {Lee},
  \citenamefont {Smith}, \citenamefont {Pagano}, \citenamefont {Potirniche},
  \citenamefont {Potter}, \citenamefont {Vishwanath} \emph
  {et~al.}}]{zhang2017discrete}%
  \BibitemOpen
  \bibfield  {author} {\bibinfo {author} {\bibfnamefont {J.}~\bibnamefont
  {Zhang}}, \bibinfo {author} {\bibfnamefont {P.}~\bibnamefont {Hess}},
  \bibinfo {author} {\bibfnamefont {A.}~\bibnamefont {Kyprianidis}}, \bibinfo
  {author} {\bibfnamefont {P.}~\bibnamefont {Becker}}, \bibinfo {author}
  {\bibfnamefont {A.}~\bibnamefont {Lee}}, \bibinfo {author} {\bibfnamefont
  {J.}~\bibnamefont {Smith}}, \bibinfo {author} {\bibfnamefont
  {G.}~\bibnamefont {Pagano}}, \bibinfo {author} {\bibfnamefont {I.-D.}\
  \bibnamefont {Potirniche}}, \bibinfo {author} {\bibfnamefont {A.~C.}\
  \bibnamefont {Potter}}, \bibinfo {author} {\bibfnamefont {A.}~\bibnamefont
  {Vishwanath}},  \emph {et~al.},\ }\href@noop {} {\bibfield  {journal}
  {\bibinfo  {journal} {Nature}\ }\textbf {\bibinfo {volume} {543}},\ \bibinfo
  {pages} {217} (\bibinfo {year} {2017}{\natexlab{b}})}\BibitemShut {NoStop}%
\bibitem [{\citenamefont {Zhang}\ \emph {et~al.}(2018)\citenamefont {Zhang},
  \citenamefont {Zhang}, \citenamefont {Shen}, \citenamefont {Zhang},
  \citenamefont {Zhang}, \citenamefont {Yung}, \citenamefont {Casanova},
  \citenamefont {Pedernales}, \citenamefont {Lamata}, \citenamefont {Solano},\
  and\ \citenamefont {Kim}}]{Zhang2018}%
  \BibitemOpen
  \bibfield  {author} {\bibinfo {author} {\bibfnamefont {X.}~\bibnamefont
  {Zhang}}, \bibinfo {author} {\bibfnamefont {K.}~\bibnamefont {Zhang}},
  \bibinfo {author} {\bibfnamefont {Y.}~\bibnamefont {Shen}}, \bibinfo {author}
  {\bibfnamefont {S.}~\bibnamefont {Zhang}}, \bibinfo {author} {\bibfnamefont
  {J.-N.}\ \bibnamefont {Zhang}}, \bibinfo {author} {\bibfnamefont {M.-H.}\
  \bibnamefont {Yung}}, \bibinfo {author} {\bibfnamefont {J.}~\bibnamefont
  {Casanova}}, \bibinfo {author} {\bibfnamefont {J.~S.}\ \bibnamefont
  {Pedernales}}, \bibinfo {author} {\bibfnamefont {L.}~\bibnamefont {Lamata}},
  \bibinfo {author} {\bibfnamefont {E.}~\bibnamefont {Solano}}, \ and\ \bibinfo
  {author} {\bibfnamefont {K.}~\bibnamefont {Kim}},\ }\href {\doibase
  10.1038/s41467-017-02507-y} {\bibfield  {journal} {\bibinfo  {journal}
  {Nature Communications}\ }\textbf {\bibinfo {volume} {9}} (\bibinfo {year}
  {2018}),\ 10.1038/s41467-017-02507-y}\BibitemShut {NoStop}%
\bibitem [{\citenamefont {Gorman}\ \emph {et~al.}(2018)\citenamefont {Gorman},
  \citenamefont {Hemmerling}, \citenamefont {Megidish}, \citenamefont
  {Moeller}, \citenamefont {Schindler}, \citenamefont {Sarovar},\ and\
  \citenamefont {Haeffner}}]{Gorman2018}%
  \BibitemOpen
  \bibfield  {author} {\bibinfo {author} {\bibfnamefont {D.~J.}\ \bibnamefont
  {Gorman}}, \bibinfo {author} {\bibfnamefont {B.}~\bibnamefont {Hemmerling}},
  \bibinfo {author} {\bibfnamefont {E.}~\bibnamefont {Megidish}}, \bibinfo
  {author} {\bibfnamefont {S.~A.}\ \bibnamefont {Moeller}}, \bibinfo {author}
  {\bibfnamefont {P.}~\bibnamefont {Schindler}}, \bibinfo {author}
  {\bibfnamefont {M.}~\bibnamefont {Sarovar}}, \ and\ \bibinfo {author}
  {\bibfnamefont {H.}~\bibnamefont {Haeffner}},\ }\href {\doibase
  10.1103/PhysRevX.8.011038} {\bibfield  {journal} {\bibinfo  {journal} {Phys.
  Rev. X}\ }\textbf {\bibinfo {volume} {8}},\ \bibinfo {pages} {011038}
  (\bibinfo {year} {2018})}\BibitemShut {NoStop}%
\bibitem [{\citenamefont {Barreiro}\ \emph {et~al.}(2011)\citenamefont
  {Barreiro}, \citenamefont {Müller}, \citenamefont {Schindler}, \citenamefont
  {Nigg}, \citenamefont {Monz}, \citenamefont {Chwalla}, \citenamefont
  {Hennrich}, \citenamefont {Roos}, \citenamefont {Zoller},\ and\ \citenamefont
  {Blatt}}]{Barreiro2011}%
  \BibitemOpen
  \bibfield  {author} {\bibinfo {author} {\bibfnamefont {J.~T.}\ \bibnamefont
  {Barreiro}}, \bibinfo {author} {\bibfnamefont {M.}~\bibnamefont {Müller}},
  \bibinfo {author} {\bibfnamefont {P.}~\bibnamefont {Schindler}}, \bibinfo
  {author} {\bibfnamefont {D.}~\bibnamefont {Nigg}}, \bibinfo {author}
  {\bibfnamefont {T.}~\bibnamefont {Monz}}, \bibinfo {author} {\bibfnamefont
  {M.}~\bibnamefont {Chwalla}}, \bibinfo {author} {\bibfnamefont
  {M.}~\bibnamefont {Hennrich}}, \bibinfo {author} {\bibfnamefont {C.~F.}\
  \bibnamefont {Roos}}, \bibinfo {author} {\bibfnamefont {P.}~\bibnamefont
  {Zoller}}, \ and\ \bibinfo {author} {\bibfnamefont {R.}~\bibnamefont
  {Blatt}},\ }\href {\doibase 10.1038/nature09801} {\bibfield  {journal}
  {\bibinfo  {journal} {Nature}\ }\textbf {\bibinfo {volume} {470}},\ \bibinfo
  {pages} {486} (\bibinfo {year} {2011})}\BibitemShut {NoStop}%
\bibitem [{\citenamefont {Lanyon}\ \emph {et~al.}(2011)\citenamefont {Lanyon},
  \citenamefont {Hempel}, \citenamefont {Nigg}, \citenamefont {M{\"u}ller},
  \citenamefont {Gerritsma}, \citenamefont {Z{\"a}hringer}, \citenamefont
  {Schindler}, \citenamefont {Barreiro}, \citenamefont {Rambach}, \citenamefont
  {Kirchmair} \emph {et~al.}}]{lanyon2011universal}%
  \BibitemOpen
  \bibfield  {author} {\bibinfo {author} {\bibfnamefont {B.~P.}\ \bibnamefont
  {Lanyon}}, \bibinfo {author} {\bibfnamefont {C.}~\bibnamefont {Hempel}},
  \bibinfo {author} {\bibfnamefont {D.}~\bibnamefont {Nigg}}, \bibinfo {author}
  {\bibfnamefont {M.}~\bibnamefont {M{\"u}ller}}, \bibinfo {author}
  {\bibfnamefont {R.}~\bibnamefont {Gerritsma}}, \bibinfo {author}
  {\bibfnamefont {F.}~\bibnamefont {Z{\"a}hringer}}, \bibinfo {author}
  {\bibfnamefont {P.}~\bibnamefont {Schindler}}, \bibinfo {author}
  {\bibfnamefont {J.~T.}\ \bibnamefont {Barreiro}}, \bibinfo {author}
  {\bibfnamefont {M.}~\bibnamefont {Rambach}}, \bibinfo {author} {\bibfnamefont
  {G.}~\bibnamefont {Kirchmair}},  \emph {et~al.},\ }\href@noop {} {\bibfield
  {journal} {\bibinfo  {journal} {Science}\ }\textbf {\bibinfo {volume}
  {334}},\ \bibinfo {pages} {57} (\bibinfo {year} {2011})}\BibitemShut
  {NoStop}%
\bibitem [{\citenamefont {Schindler}\ \emph {et~al.}(2013)\citenamefont
  {Schindler}, \citenamefont {M\"{u}ller}, \citenamefont {Nigg}, \citenamefont
  {Barreiro}, \citenamefont {Martinez}, \citenamefont {Hennrich}, \citenamefont
  {Monz}, \citenamefont {Diehl}, \citenamefont {Zoller},\ and\ \citenamefont
  {Blatt}}]{Schindler2013}%
  \BibitemOpen
  \bibfield  {author} {\bibinfo {author} {\bibfnamefont {P.}~\bibnamefont
  {Schindler}}, \bibinfo {author} {\bibfnamefont {M.}~\bibnamefont
  {M\"{u}ller}}, \bibinfo {author} {\bibfnamefont {D.}~\bibnamefont {Nigg}},
  \bibinfo {author} {\bibfnamefont {J.~T.}\ \bibnamefont {Barreiro}}, \bibinfo
  {author} {\bibfnamefont {E.~A.}\ \bibnamefont {Martinez}}, \bibinfo {author}
  {\bibfnamefont {M.}~\bibnamefont {Hennrich}}, \bibinfo {author}
  {\bibfnamefont {T.}~\bibnamefont {Monz}}, \bibinfo {author} {\bibfnamefont
  {S.}~\bibnamefont {Diehl}}, \bibinfo {author} {\bibfnamefont
  {P.}~\bibnamefont {Zoller}}, \ and\ \bibinfo {author} {\bibfnamefont
  {R.}~\bibnamefont {Blatt}},\ }\href {\doibase 10.1038/nphys2630} {\bibfield
  {journal} {\bibinfo  {journal} {Nature Physics}\ }\textbf {\bibinfo {volume}
  {9}},\ \bibinfo {pages} {361} (\bibinfo {year} {2013})}\BibitemShut {NoStop}%
\bibitem [{\citenamefont {Martinez}\ \emph {et~al.}(2016)\citenamefont
  {Martinez}, \citenamefont {Muschik}, \citenamefont {Schindler}, \citenamefont
  {Nigg}, \citenamefont {Erhard}, \citenamefont {Heyl}, \citenamefont {Hauke},
  \citenamefont {Dalmonte}, \citenamefont {Monz}, \citenamefont {Zoller},\ and\
  \citenamefont {Blatt}}]{Martinez2016}%
  \BibitemOpen
  \bibfield  {author} {\bibinfo {author} {\bibfnamefont {E.~A.}\ \bibnamefont
  {Martinez}}, \bibinfo {author} {\bibfnamefont {C.~A.}\ \bibnamefont
  {Muschik}}, \bibinfo {author} {\bibfnamefont {P.}~\bibnamefont {Schindler}},
  \bibinfo {author} {\bibfnamefont {D.}~\bibnamefont {Nigg}}, \bibinfo {author}
  {\bibfnamefont {A.}~\bibnamefont {Erhard}}, \bibinfo {author} {\bibfnamefont
  {M.}~\bibnamefont {Heyl}}, \bibinfo {author} {\bibfnamefont {P.}~\bibnamefont
  {Hauke}}, \bibinfo {author} {\bibfnamefont {M.}~\bibnamefont {Dalmonte}},
  \bibinfo {author} {\bibfnamefont {T.}~\bibnamefont {Monz}}, \bibinfo {author}
  {\bibfnamefont {P.}~\bibnamefont {Zoller}}, \ and\ \bibinfo {author}
  {\bibfnamefont {R.}~\bibnamefont {Blatt}},\ }\href {\doibase
  10.1038/nature18318} {\bibfield  {journal} {\bibinfo  {journal} {Nature}\
  }\textbf {\bibinfo {volume} {534}},\ \bibinfo {pages} {516} (\bibinfo {year}
  {2016})}\BibitemShut {NoStop}%
\bibitem [{\citenamefont {Hempel}\ \emph {et~al.}(2018)\citenamefont {Hempel},
  \citenamefont {Maier}, \citenamefont {Romero}, \citenamefont {McClean},
  \citenamefont {Monz}, \citenamefont {Shen}, \citenamefont {Jurcevic},
  \citenamefont {Lanyon}, \citenamefont {Love}, \citenamefont {Babbush},
  \citenamefont {Aspuru-Guzik}, \citenamefont {Blatt},\ and\ \citenamefont
  {Roos}}]{Hempel2018}%
  \BibitemOpen
  \bibfield  {author} {\bibinfo {author} {\bibfnamefont {C.}~\bibnamefont
  {Hempel}}, \bibinfo {author} {\bibfnamefont {C.}~\bibnamefont {Maier}},
  \bibinfo {author} {\bibfnamefont {J.}~\bibnamefont {Romero}}, \bibinfo
  {author} {\bibfnamefont {J.}~\bibnamefont {McClean}}, \bibinfo {author}
  {\bibfnamefont {T.}~\bibnamefont {Monz}}, \bibinfo {author} {\bibfnamefont
  {H.}~\bibnamefont {Shen}}, \bibinfo {author} {\bibfnamefont {P.}~\bibnamefont
  {Jurcevic}}, \bibinfo {author} {\bibfnamefont {B.~P.}\ \bibnamefont
  {Lanyon}}, \bibinfo {author} {\bibfnamefont {P.}~\bibnamefont {Love}},
  \bibinfo {author} {\bibfnamefont {R.}~\bibnamefont {Babbush}}, \bibinfo
  {author} {\bibfnamefont {A.}~\bibnamefont {Aspuru-Guzik}}, \bibinfo {author}
  {\bibfnamefont {R.}~\bibnamefont {Blatt}}, \ and\ \bibinfo {author}
  {\bibfnamefont {C.~F.}\ \bibnamefont {Roos}},\ }\href {\doibase
  10.1103/PhysRevX.8.031022} {\bibfield  {journal} {\bibinfo  {journal} {Phys.
  Rev. X}\ }\textbf {\bibinfo {volume} {8}},\ \bibinfo {pages} {031022}
  (\bibinfo {year} {2018})}\BibitemShut {NoStop}%
\bibitem [{\citenamefont {M{\o}lmer}\ and\ \citenamefont
  {S{\o}rensen}(1999)}]{molmer1999multiparticle}%
  \BibitemOpen
  \bibfield  {author} {\bibinfo {author} {\bibfnamefont {K.}~\bibnamefont
  {M{\o}lmer}}\ and\ \bibinfo {author} {\bibfnamefont {A.}~\bibnamefont
  {S{\o}rensen}},\ }\href@noop {} {\bibfield  {journal} {\bibinfo  {journal}
  {Physical Review Letters}\ }\textbf {\bibinfo {volume} {82}},\ \bibinfo
  {pages} {1835} (\bibinfo {year} {1999})}\BibitemShut {NoStop}%
\bibitem [{\citenamefont {S{\o}rensen}\ and\ \citenamefont
  {M{\o}lmer}(2000)}]{sorensen2000entanglement}%
  \BibitemOpen
  \bibfield  {author} {\bibinfo {author} {\bibfnamefont {A.}~\bibnamefont
  {S{\o}rensen}}\ and\ \bibinfo {author} {\bibfnamefont {K.}~\bibnamefont
  {M{\o}lmer}},\ }\href@noop {} {\bibfield  {journal} {\bibinfo  {journal}
  {Physical Review A}\ }\textbf {\bibinfo {volume} {62}},\ \bibinfo {pages}
  {022311} (\bibinfo {year} {2000})}\BibitemShut {NoStop}%
\bibitem [{\citenamefont {Roos}(2008)}]{roos2008}%
  \BibitemOpen
  \bibfield  {author} {\bibinfo {author} {\bibfnamefont {C.~F.}\ \bibnamefont
  {Roos}},\ }\href {\doibase 10.1088/1367-2630/10/1/013002} {\bibfield
  {journal} {\bibinfo  {journal} {New Journal of Physics}\ }\textbf {\bibinfo
  {volume} {10}},\ \bibinfo {pages} {013002} (\bibinfo {year}
  {2008})}\BibitemShut {NoStop}%
\bibitem [{\citenamefont {Porras}\ and\ \citenamefont
  {Cirac}(2004)}]{porras2004effective}%
  \BibitemOpen
  \bibfield  {author} {\bibinfo {author} {\bibfnamefont {D.}~\bibnamefont
  {Porras}}\ and\ \bibinfo {author} {\bibfnamefont {J.~I.}\ \bibnamefont
  {Cirac}},\ }\href@noop {} {\bibfield  {journal} {\bibinfo  {journal}
  {Physical Review Letters}\ }\textbf {\bibinfo {volume} {92}},\ \bibinfo
  {pages} {207901} (\bibinfo {year} {2004})}\BibitemShut {NoStop}%
\bibitem [{\citenamefont {Korenblit}\ \emph {et~al.}(2012)\citenamefont
  {Korenblit}, \citenamefont {Kafri}, \citenamefont {Campbell}, \citenamefont
  {Islam}, \citenamefont {Edwards}, \citenamefont {Gong}, \citenamefont {Lin},
  \citenamefont {Duan}, \citenamefont {Kim}, \citenamefont {Kim},\ and\
  \citenamefont {Monroe}}]{Korenblit2012}%
  \BibitemOpen
  \bibfield  {author} {\bibinfo {author} {\bibfnamefont {S.}~\bibnamefont
  {Korenblit}}, \bibinfo {author} {\bibfnamefont {D.}~\bibnamefont {Kafri}},
  \bibinfo {author} {\bibfnamefont {W.~C.}\ \bibnamefont {Campbell}}, \bibinfo
  {author} {\bibfnamefont {R.}~\bibnamefont {Islam}}, \bibinfo {author}
  {\bibfnamefont {E.~E.}\ \bibnamefont {Edwards}}, \bibinfo {author}
  {\bibfnamefont {Z.-X.}\ \bibnamefont {Gong}}, \bibinfo {author}
  {\bibfnamefont {G.-D.}\ \bibnamefont {Lin}}, \bibinfo {author} {\bibfnamefont
  {L.-M.}\ \bibnamefont {Duan}}, \bibinfo {author} {\bibfnamefont
  {J.}~\bibnamefont {Kim}}, \bibinfo {author} {\bibfnamefont {K.}~\bibnamefont
  {Kim}}, \ and\ \bibinfo {author} {\bibfnamefont {C.}~\bibnamefont {Monroe}},\
  }\href {\doibase 10.1088/1367-2630/14/9/095024} {\bibfield  {journal}
  {\bibinfo  {journal} {New Journal of Physics}\ }\textbf {\bibinfo {volume}
  {14}},\ \bibinfo {pages} {095024} (\bibinfo {year} {2012})}\BibitemShut
  {NoStop}%
\bibitem [{\citenamefont {Rajabi}\ \emph {et~al.}(2019)\citenamefont {Rajabi},
  \citenamefont {Motlakunta}, \citenamefont {Shih}, \citenamefont
  {Kotibhaskar}, \citenamefont {Quraishi}, \citenamefont {Ajoy},\ and\
  \citenamefont {Islam}}]{rajabi2019dynamical}%
  \BibitemOpen
  \bibfield  {author} {\bibinfo {author} {\bibfnamefont {F.}~\bibnamefont
  {Rajabi}}, \bibinfo {author} {\bibfnamefont {S.}~\bibnamefont {Motlakunta}},
  \bibinfo {author} {\bibfnamefont {C.-Y.}\ \bibnamefont {Shih}}, \bibinfo
  {author} {\bibfnamefont {N.}~\bibnamefont {Kotibhaskar}}, \bibinfo {author}
  {\bibfnamefont {Q.}~\bibnamefont {Quraishi}}, \bibinfo {author}
  {\bibfnamefont {A.}~\bibnamefont {Ajoy}}, \ and\ \bibinfo {author}
  {\bibfnamefont {R.}~\bibnamefont {Islam}},\ }\href@noop {} {\bibfield
  {journal} {\bibinfo  {journal} {npj Quantum Information}\ }\textbf {\bibinfo
  {volume} {5}},\ \bibinfo {pages} {1} (\bibinfo {year} {2019})}\BibitemShut
  {NoStop}%
\bibitem [{\citenamefont {Shapira}\ \emph {et~al.}(2019)\citenamefont
  {Shapira}, \citenamefont {Shaniv}, \citenamefont {Manovitz}, \citenamefont
  {Akerman}, \citenamefont {Peleg}, \citenamefont {Gazit}, \citenamefont
  {Ozeri},\ and\ \citenamefont {Stern}}]{shapira2019theory}%
  \BibitemOpen
  \bibfield  {author} {\bibinfo {author} {\bibfnamefont {Y.}~\bibnamefont
  {Shapira}}, \bibinfo {author} {\bibfnamefont {R.}~\bibnamefont {Shaniv}},
  \bibinfo {author} {\bibfnamefont {T.}~\bibnamefont {Manovitz}}, \bibinfo
  {author} {\bibfnamefont {N.}~\bibnamefont {Akerman}}, \bibinfo {author}
  {\bibfnamefont {L.}~\bibnamefont {Peleg}}, \bibinfo {author} {\bibfnamefont
  {L.}~\bibnamefont {Gazit}}, \bibinfo {author} {\bibfnamefont
  {R.}~\bibnamefont {Ozeri}}, \ and\ \bibinfo {author} {\bibfnamefont
  {A.}~\bibnamefont {Stern}},\ }\href@noop {} {\bibfield  {journal} {\bibinfo
  {journal} {Phys. Rev. A}\ }\textbf {\bibinfo {volume} {101}},\ \bibinfo
  {pages} {032330} (\bibinfo {year} {2019})}\BibitemShut {NoStop}%
\bibitem [{\citenamefont {Davoudi}\ \emph {et~al.}(2020)\citenamefont
  {Davoudi}, \citenamefont {Hafezi}, \citenamefont {Monroe}, \citenamefont
  {Pagano}, \citenamefont {Seif},\ and\ \citenamefont
  {Shaw}}]{davoudi2019towards}%
  \BibitemOpen
  \bibfield  {author} {\bibinfo {author} {\bibfnamefont {Z.}~\bibnamefont
  {Davoudi}}, \bibinfo {author} {\bibfnamefont {M.}~\bibnamefont {Hafezi}},
  \bibinfo {author} {\bibfnamefont {C.}~\bibnamefont {Monroe}}, \bibinfo
  {author} {\bibfnamefont {G.}~\bibnamefont {Pagano}}, \bibinfo {author}
  {\bibfnamefont {A.}~\bibnamefont {Seif}}, \ and\ \bibinfo {author}
  {\bibfnamefont {A.}~\bibnamefont {Shaw}},\ }\href {\doibase
  10.1103/PhysRevResearch.2.023015} {\bibfield  {journal} {\bibinfo  {journal}
  {Phys. Rev. Research}\ }\textbf {\bibinfo {volume} {2}},\ \bibinfo {pages}
  {023015} (\bibinfo {year} {2020})}\BibitemShut {NoStop}%
\bibitem [{\citenamefont {Lu}\ \emph {et~al.}(2019)\citenamefont {Lu},
  \citenamefont {Zhang}, \citenamefont {Zhang}, \citenamefont {Chen},
  \citenamefont {Shen}, \citenamefont {Zhang}, \citenamefont {Zhang},\ and\
  \citenamefont {Kim}}]{lu2019global}%
  \BibitemOpen
  \bibfield  {author} {\bibinfo {author} {\bibfnamefont {Y.}~\bibnamefont
  {Lu}}, \bibinfo {author} {\bibfnamefont {S.}~\bibnamefont {Zhang}}, \bibinfo
  {author} {\bibfnamefont {K.}~\bibnamefont {Zhang}}, \bibinfo {author}
  {\bibfnamefont {W.}~\bibnamefont {Chen}}, \bibinfo {author} {\bibfnamefont
  {Y.}~\bibnamefont {Shen}}, \bibinfo {author} {\bibfnamefont {J.}~\bibnamefont
  {Zhang}}, \bibinfo {author} {\bibfnamefont {J.-N.}\ \bibnamefont {Zhang}}, \
  and\ \bibinfo {author} {\bibfnamefont {K.}~\bibnamefont {Kim}},\ }\href@noop
  {} {\bibfield  {journal} {\bibinfo  {journal} {Nature}\ }\textbf {\bibinfo
  {volume} {572}},\ \bibinfo {pages} {363} (\bibinfo {year}
  {2019})}\BibitemShut {NoStop}%
\bibitem [{\citenamefont {Figgatt}\ \emph {et~al.}(2019)\citenamefont
  {Figgatt}, \citenamefont {Ostrander}, \citenamefont {Linke}, \citenamefont
  {Landsman}, \citenamefont {Zhu}, \citenamefont {Maslov},\ and\ \citenamefont
  {Monroe}}]{figgatt2019parallel}%
  \BibitemOpen
  \bibfield  {author} {\bibinfo {author} {\bibfnamefont {C.}~\bibnamefont
  {Figgatt}}, \bibinfo {author} {\bibfnamefont {A.}~\bibnamefont {Ostrander}},
  \bibinfo {author} {\bibfnamefont {N.~M.}\ \bibnamefont {Linke}}, \bibinfo
  {author} {\bibfnamefont {K.~A.}\ \bibnamefont {Landsman}}, \bibinfo {author}
  {\bibfnamefont {D.}~\bibnamefont {Zhu}}, \bibinfo {author} {\bibfnamefont
  {D.}~\bibnamefont {Maslov}}, \ and\ \bibinfo {author} {\bibfnamefont
  {C.}~\bibnamefont {Monroe}},\ }\href@noop {} {\bibfield  {journal} {\bibinfo
  {journal} {Nature}\ }\textbf {\bibinfo {volume} {572}},\ \bibinfo {pages}
  {368} (\bibinfo {year} {2019})}\BibitemShut {NoStop}%
\bibitem [{\citenamefont {Hofstadter}(1976)}]{hofstader1976butterfly}%
  \BibitemOpen
  \bibfield  {author} {\bibinfo {author} {\bibfnamefont {D.~R.}\ \bibnamefont
  {Hofstadter}},\ }\href@noop {} {\bibfield  {journal} {\bibinfo  {journal}
  {Phys. Rev. B}\ }\textbf {\bibinfo {volume} {14}},\ \bibinfo {pages} {2239}
  (\bibinfo {year} {1976})}\BibitemShut {NoStop}%
\bibitem [{\citenamefont {D.~J.~Thouless}\ and\ \citenamefont {den
  Nijs}(1982)}]{tknn1982iqhe}%
  \BibitemOpen
  \bibfield  {author} {\bibinfo {author} {\bibfnamefont {M.~P.~N.}\
  \bibnamefont {D.~J.~Thouless}, \bibfnamefont {M.~Kohmoto}}\ and\ \bibinfo
  {author} {\bibfnamefont {M.}~\bibnamefont {den Nijs}},\ }\href@noop {}
  {\bibfield  {journal} {\bibinfo  {journal} {Phys. Rev. Lett.}\ }\textbf
  {\bibinfo {volume} {49}},\ \bibinfo {pages} {405} (\bibinfo {year}
  {1982})}\BibitemShut {NoStop}%
\bibitem [{\citenamefont {Jaksch}\ and\ \citenamefont
  {Zoller}(2003)}]{jaksch2003creation}%
  \BibitemOpen
  \bibfield  {author} {\bibinfo {author} {\bibfnamefont {D.}~\bibnamefont
  {Jaksch}}\ and\ \bibinfo {author} {\bibfnamefont {P.}~\bibnamefont
  {Zoller}},\ }\href@noop {} {\bibfield  {journal} {\bibinfo  {journal} {New
  Journal of Physics}\ }\textbf {\bibinfo {volume} {5}},\ \bibinfo {pages} {56}
  (\bibinfo {year} {2003})}\BibitemShut {NoStop}%
\bibitem [{\citenamefont {Miyake}\ \emph {et~al.}(2013)\citenamefont {Miyake},
  \citenamefont {Siviloglou}, \citenamefont {Kennedy}, \citenamefont {Burton},\
  and\ \citenamefont {Ketterle}}]{miyake2013realizing}%
  \BibitemOpen
  \bibfield  {author} {\bibinfo {author} {\bibfnamefont {H.}~\bibnamefont
  {Miyake}}, \bibinfo {author} {\bibfnamefont {G.~A.}\ \bibnamefont
  {Siviloglou}}, \bibinfo {author} {\bibfnamefont {C.~J.}\ \bibnamefont
  {Kennedy}}, \bibinfo {author} {\bibfnamefont {W.~C.}\ \bibnamefont {Burton}},
  \ and\ \bibinfo {author} {\bibfnamefont {W.}~\bibnamefont {Ketterle}},\
  }\href@noop {} {\bibfield  {journal} {\bibinfo  {journal} {Physical Review
  Letters}\ }\textbf {\bibinfo {volume} {111}},\ \bibinfo {pages} {185302}
  (\bibinfo {year} {2013})}\BibitemShut {NoStop}%
\bibitem [{\citenamefont {Dalibard}\ \emph {et~al.}(2011)\citenamefont
  {Dalibard}, \citenamefont {Gerbier}, \citenamefont {Juzeli{\=u}nas},\ and\
  \citenamefont {{\"O}hberg}}]{dalibard2011colloquium}%
  \BibitemOpen
  \bibfield  {author} {\bibinfo {author} {\bibfnamefont {J.}~\bibnamefont
  {Dalibard}}, \bibinfo {author} {\bibfnamefont {F.}~\bibnamefont {Gerbier}},
  \bibinfo {author} {\bibfnamefont {G.}~\bibnamefont {Juzeli{\=u}nas}}, \ and\
  \bibinfo {author} {\bibfnamefont {P.}~\bibnamefont {{\"O}hberg}},\
  }\href@noop {} {\bibfield  {journal} {\bibinfo  {journal} {Reviews of Modern
  Physics}\ }\textbf {\bibinfo {volume} {83}},\ \bibinfo {pages} {1523}
  (\bibinfo {year} {2011})}\BibitemShut {NoStop}%
\bibitem [{\citenamefont {Bloch}\ \emph {et~al.}(2012)\citenamefont {Bloch},
  \citenamefont {Dalibard},\ and\ \citenamefont
  {Nascimbene}}]{bloch2012quantum}%
  \BibitemOpen
  \bibfield  {author} {\bibinfo {author} {\bibfnamefont {I.}~\bibnamefont
  {Bloch}}, \bibinfo {author} {\bibfnamefont {J.}~\bibnamefont {Dalibard}}, \
  and\ \bibinfo {author} {\bibfnamefont {S.}~\bibnamefont {Nascimbene}},\
  }\href@noop {} {\bibfield  {journal} {\bibinfo  {journal} {Nature Physics}\
  }\textbf {\bibinfo {volume} {8}},\ \bibinfo {pages} {267} (\bibinfo {year}
  {2012})}\BibitemShut {NoStop}%
\bibitem [{\citenamefont {Lin}\ \emph {et~al.}(2011)\citenamefont {Lin},
  \citenamefont {Jim{\'e}nez-Garc{\'\i}a},\ and\ \citenamefont
  {Spielman}}]{lin2011spin}%
  \BibitemOpen
  \bibfield  {author} {\bibinfo {author} {\bibfnamefont {Y.-J.}\ \bibnamefont
  {Lin}}, \bibinfo {author} {\bibfnamefont {K.}~\bibnamefont
  {Jim{\'e}nez-Garc{\'\i}a}}, \ and\ \bibinfo {author} {\bibfnamefont {I.~B.}\
  \bibnamefont {Spielman}},\ }\href@noop {} {\bibfield  {journal} {\bibinfo
  {journal} {Nature}\ }\textbf {\bibinfo {volume} {471}},\ \bibinfo {pages}
  {83} (\bibinfo {year} {2011})}\BibitemShut {NoStop}%
\bibitem [{\citenamefont {Struck}\ \emph {et~al.}(2013)\citenamefont {Struck},
  \citenamefont {Weinberg}, \citenamefont {{\"O}lschl{\"a}ger}, \citenamefont
  {Windpassinger}, \citenamefont {Simonet}, \citenamefont {Sengstock},
  \citenamefont {H{\"o}ppner}, \citenamefont {Hauke}, \citenamefont {Eckardt},
  \citenamefont {Lewenstein} \emph {et~al.}}]{struck2013engineering}%
  \BibitemOpen
  \bibfield  {author} {\bibinfo {author} {\bibfnamefont {J.}~\bibnamefont
  {Struck}}, \bibinfo {author} {\bibfnamefont {M.}~\bibnamefont {Weinberg}},
  \bibinfo {author} {\bibfnamefont {C.}~\bibnamefont {{\"O}lschl{\"a}ger}},
  \bibinfo {author} {\bibfnamefont {P.}~\bibnamefont {Windpassinger}}, \bibinfo
  {author} {\bibfnamefont {J.}~\bibnamefont {Simonet}}, \bibinfo {author}
  {\bibfnamefont {K.}~\bibnamefont {Sengstock}}, \bibinfo {author}
  {\bibfnamefont {R.}~\bibnamefont {H{\"o}ppner}}, \bibinfo {author}
  {\bibfnamefont {P.}~\bibnamefont {Hauke}}, \bibinfo {author} {\bibfnamefont
  {A.}~\bibnamefont {Eckardt}}, \bibinfo {author} {\bibfnamefont
  {M.}~\bibnamefont {Lewenstein}},  \emph {et~al.},\ }\href@noop {} {\bibfield
  {journal} {\bibinfo  {journal} {Nature Physics}\ }\textbf {\bibinfo {volume}
  {9}},\ \bibinfo {pages} {738} (\bibinfo {year} {2013})}\BibitemShut {NoStop}%
\bibitem [{\citenamefont {Lin}\ \emph {et~al.}(2009{\natexlab{a}})\citenamefont
  {Lin}, \citenamefont {Compton}, \citenamefont {Jim{\'e}nez-Garc{\'\i}a},
  \citenamefont {Porto},\ and\ \citenamefont {Spielman}}]{lin2009synthetic}%
  \BibitemOpen
  \bibfield  {author} {\bibinfo {author} {\bibfnamefont {Y.-J.}\ \bibnamefont
  {Lin}}, \bibinfo {author} {\bibfnamefont {R.~L.}\ \bibnamefont {Compton}},
  \bibinfo {author} {\bibfnamefont {K.}~\bibnamefont
  {Jim{\'e}nez-Garc{\'\i}a}}, \bibinfo {author} {\bibfnamefont {J.~V.}\
  \bibnamefont {Porto}}, \ and\ \bibinfo {author} {\bibfnamefont {I.~B.}\
  \bibnamefont {Spielman}},\ }\href@noop {} {\bibfield  {journal} {\bibinfo
  {journal} {Nature}\ }\textbf {\bibinfo {volume} {462}},\ \bibinfo {pages}
  {628} (\bibinfo {year} {2009}{\natexlab{a}})}\BibitemShut {NoStop}%
\bibitem [{\citenamefont {Aidelsburger}\ \emph {et~al.}(2013)\citenamefont
  {Aidelsburger}, \citenamefont {Atala}, \citenamefont {Lohse}, \citenamefont
  {Barreiro}, \citenamefont {Paredes},\ and\ \citenamefont
  {Bloch}}]{aidelsburger2013realization}%
  \BibitemOpen
  \bibfield  {author} {\bibinfo {author} {\bibfnamefont {M.}~\bibnamefont
  {Aidelsburger}}, \bibinfo {author} {\bibfnamefont {M.}~\bibnamefont {Atala}},
  \bibinfo {author} {\bibfnamefont {M.}~\bibnamefont {Lohse}}, \bibinfo
  {author} {\bibfnamefont {J.~T.}\ \bibnamefont {Barreiro}}, \bibinfo {author}
  {\bibfnamefont {B.}~\bibnamefont {Paredes}}, \ and\ \bibinfo {author}
  {\bibfnamefont {I.}~\bibnamefont {Bloch}},\ }\href@noop {} {\bibfield
  {journal} {\bibinfo  {journal} {Physical Review Letters}\ }\textbf {\bibinfo
  {volume} {111}},\ \bibinfo {pages} {185301} (\bibinfo {year}
  {2013})}\BibitemShut {NoStop}%
\bibitem [{\citenamefont {Mancini}\ \emph {et~al.}(2015)\citenamefont
  {Mancini}, \citenamefont {Pagano}, \citenamefont {Cappellini}, \citenamefont
  {Livi}, \citenamefont {Rider}, \citenamefont {Catani}, \citenamefont {Sias},
  \citenamefont {Zoller}, \citenamefont {Inguscio}, \citenamefont {Dalmonte}
  \emph {et~al.}}]{mancini2015observation}%
  \BibitemOpen
  \bibfield  {author} {\bibinfo {author} {\bibfnamefont {M.}~\bibnamefont
  {Mancini}}, \bibinfo {author} {\bibfnamefont {G.}~\bibnamefont {Pagano}},
  \bibinfo {author} {\bibfnamefont {G.}~\bibnamefont {Cappellini}}, \bibinfo
  {author} {\bibfnamefont {L.}~\bibnamefont {Livi}}, \bibinfo {author}
  {\bibfnamefont {M.}~\bibnamefont {Rider}}, \bibinfo {author} {\bibfnamefont
  {J.}~\bibnamefont {Catani}}, \bibinfo {author} {\bibfnamefont
  {C.}~\bibnamefont {Sias}}, \bibinfo {author} {\bibfnamefont {P.}~\bibnamefont
  {Zoller}}, \bibinfo {author} {\bibfnamefont {M.}~\bibnamefont {Inguscio}},
  \bibinfo {author} {\bibfnamefont {M.}~\bibnamefont {Dalmonte}},  \emph
  {et~al.},\ }\href@noop {} {\bibfield  {journal} {\bibinfo  {journal}
  {Science}\ }\textbf {\bibinfo {volume} {349}},\ \bibinfo {pages} {1510}
  (\bibinfo {year} {2015})}\BibitemShut {NoStop}%
\bibitem [{\citenamefont {Gaebler}\ \emph {et~al.}(2016)\citenamefont
  {Gaebler}, \citenamefont {Tan}, \citenamefont {Lin}, \citenamefont {Wan},
  \citenamefont {Bowler}, \citenamefont {Keith}, \citenamefont {Glancy},
  \citenamefont {Coakley}, \citenamefont {Knill}, \citenamefont {Leibfried},\
  and\ \citenamefont {Wineland}}]{gaebler2016}%
  \BibitemOpen
  \bibfield  {author} {\bibinfo {author} {\bibfnamefont {J.~P.}\ \bibnamefont
  {Gaebler}}, \bibinfo {author} {\bibfnamefont {T.~R.}\ \bibnamefont {Tan}},
  \bibinfo {author} {\bibfnamefont {Y.}~\bibnamefont {Lin}}, \bibinfo {author}
  {\bibfnamefont {Y.}~\bibnamefont {Wan}}, \bibinfo {author} {\bibfnamefont
  {R.}~\bibnamefont {Bowler}}, \bibinfo {author} {\bibfnamefont {A.~C.}\
  \bibnamefont {Keith}}, \bibinfo {author} {\bibfnamefont {S.}~\bibnamefont
  {Glancy}}, \bibinfo {author} {\bibfnamefont {K.}~\bibnamefont {Coakley}},
  \bibinfo {author} {\bibfnamefont {E.}~\bibnamefont {Knill}}, \bibinfo
  {author} {\bibfnamefont {D.}~\bibnamefont {Leibfried}}, \ and\ \bibinfo
  {author} {\bibfnamefont {D.~J.}\ \bibnamefont {Wineland}},\ }\href {\doibase
  10.1103/PhysRevLett.117.060505} {\bibfield  {journal} {\bibinfo  {journal}
  {Phys. Rev. Lett.}\ }\textbf {\bibinfo {volume} {117}},\ \bibinfo {pages}
  {060505} (\bibinfo {year} {2016})}\BibitemShut {NoStop}%
\bibitem [{\citenamefont {Debnath}\ \emph {et~al.}(2016)\citenamefont
  {Debnath}, \citenamefont {Linke}, \citenamefont {Figgatt}, \citenamefont
  {Landsman}, \citenamefont {Wright},\ and\ \citenamefont
  {Monroe}}]{debnath2016demonstration}%
  \BibitemOpen
  \bibfield  {author} {\bibinfo {author} {\bibfnamefont {S.}~\bibnamefont
  {Debnath}}, \bibinfo {author} {\bibfnamefont {N.~M.}\ \bibnamefont {Linke}},
  \bibinfo {author} {\bibfnamefont {C.}~\bibnamefont {Figgatt}}, \bibinfo
  {author} {\bibfnamefont {K.~A.}\ \bibnamefont {Landsman}}, \bibinfo {author}
  {\bibfnamefont {K.}~\bibnamefont {Wright}}, \ and\ \bibinfo {author}
  {\bibfnamefont {C.}~\bibnamefont {Monroe}},\ }\href@noop {} {\bibfield
  {journal} {\bibinfo  {journal} {Nature}\ }\textbf {\bibinfo {volume} {536}},\
  \bibinfo {pages} {63} (\bibinfo {year} {2016})}\BibitemShut {NoStop}%
\bibitem [{\citenamefont {Barenco}\ \emph {et~al.}(1995)\citenamefont
  {Barenco}, \citenamefont {Bennett}, \citenamefont {Cleve}, \citenamefont
  {DiVincenzo}, \citenamefont {Margolus}, \citenamefont {Shor}, \citenamefont
  {Sleator}, \citenamefont {Smolin},\ and\ \citenamefont
  {Weinfurter}}]{barenco1995}%
  \BibitemOpen
  \bibfield  {author} {\bibinfo {author} {\bibfnamefont {A.}~\bibnamefont
  {Barenco}}, \bibinfo {author} {\bibfnamefont {C.~H.}\ \bibnamefont
  {Bennett}}, \bibinfo {author} {\bibfnamefont {R.}~\bibnamefont {Cleve}},
  \bibinfo {author} {\bibfnamefont {D.~P.}\ \bibnamefont {DiVincenzo}},
  \bibinfo {author} {\bibfnamefont {N.}~\bibnamefont {Margolus}}, \bibinfo
  {author} {\bibfnamefont {P.}~\bibnamefont {Shor}}, \bibinfo {author}
  {\bibfnamefont {T.}~\bibnamefont {Sleator}}, \bibinfo {author} {\bibfnamefont
  {J.~A.}\ \bibnamefont {Smolin}}, \ and\ \bibinfo {author} {\bibfnamefont
  {H.}~\bibnamefont {Weinfurter}},\ }\href {\doibase 10.1103/PhysRevA.52.3457}
  {\bibfield  {journal} {\bibinfo  {journal} {Phys. Rev. A}\ }\textbf {\bibinfo
  {volume} {52}},\ \bibinfo {pages} {3457} (\bibinfo {year}
  {1995})}\BibitemShut {NoStop}%
\bibitem [{\citenamefont {Preskill}(2018)}]{preskill2018quantum}%
  \BibitemOpen
  \bibfield  {author} {\bibinfo {author} {\bibfnamefont {J.}~\bibnamefont
  {Preskill}},\ }\href@noop {} {\bibfield  {journal} {\bibinfo  {journal}
  {Quantum}\ }\textbf {\bibinfo {volume} {2}},\ \bibinfo {pages} {79} (\bibinfo
  {year} {2018})}\BibitemShut {NoStop}%
\bibitem [{\citenamefont {Mintert}\ and\ \citenamefont
  {Wunderlich}(2001)}]{mintert2001ion}%
  \BibitemOpen
  \bibfield  {author} {\bibinfo {author} {\bibfnamefont {F.}~\bibnamefont
  {Mintert}}\ and\ \bibinfo {author} {\bibfnamefont {C.}~\bibnamefont
  {Wunderlich}},\ }\href {\doibase 10.1103/PhysRevLett.87.257904} {\bibfield
  {journal} {\bibinfo  {journal} {Phys. Rev. Lett.}\ }\textbf {\bibinfo
  {volume} {87}},\ \bibinfo {pages} {257904} (\bibinfo {year}
  {2001})}\BibitemShut {NoStop}%
\bibitem [{\citenamefont {Johanning}\ \emph {et~al.}(2009)\citenamefont
  {Johanning}, \citenamefont {Braun}, \citenamefont {Timoney}, \citenamefont
  {Elman}, \citenamefont {Neuhauser},\ and\ \citenamefont
  {Wunderlich}}]{johanning2009individual}%
  \BibitemOpen
  \bibfield  {author} {\bibinfo {author} {\bibfnamefont {M.}~\bibnamefont
  {Johanning}}, \bibinfo {author} {\bibfnamefont {A.}~\bibnamefont {Braun}},
  \bibinfo {author} {\bibfnamefont {N.}~\bibnamefont {Timoney}}, \bibinfo
  {author} {\bibfnamefont {V.}~\bibnamefont {Elman}}, \bibinfo {author}
  {\bibfnamefont {W.}~\bibnamefont {Neuhauser}}, \ and\ \bibinfo {author}
  {\bibfnamefont {C.}~\bibnamefont {Wunderlich}},\ }\href {\doibase
  10.1103/PhysRevLett.102.073004} {\bibfield  {journal} {\bibinfo  {journal}
  {Phys. Rev. Lett.}\ }\textbf {\bibinfo {volume} {102}},\ \bibinfo {pages}
  {073004} (\bibinfo {year} {2009})}\BibitemShut {NoStop}%
\bibitem [{\citenamefont {Timoney}\ \emph {et~al.}(2011)\citenamefont
  {Timoney}, \citenamefont {Baumgart}, \citenamefont {Johanning}, \citenamefont
  {Var{\'o}n}, \citenamefont {Plenio}, \citenamefont {Retzker},\ and\
  \citenamefont {Wunderlich}}]{timoney2011quantum}%
  \BibitemOpen
  \bibfield  {author} {\bibinfo {author} {\bibfnamefont {N.}~\bibnamefont
  {Timoney}}, \bibinfo {author} {\bibfnamefont {I.}~\bibnamefont {Baumgart}},
  \bibinfo {author} {\bibfnamefont {M.}~\bibnamefont {Johanning}}, \bibinfo
  {author} {\bibfnamefont {A.}~\bibnamefont {Var{\'o}n}}, \bibinfo {author}
  {\bibfnamefont {M.~B.}\ \bibnamefont {Plenio}}, \bibinfo {author}
  {\bibfnamefont {A.}~\bibnamefont {Retzker}}, \ and\ \bibinfo {author}
  {\bibfnamefont {C.}~\bibnamefont {Wunderlich}},\ }\href@noop {} {\bibfield
  {journal} {\bibinfo  {journal} {Nature}\ }\textbf {\bibinfo {volume} {476}},\
  \bibinfo {pages} {185} (\bibinfo {year} {2011})}\BibitemShut {NoStop}%
\bibitem [{\citenamefont {Gra\ss{}}\ \emph {et~al.}(2015)\citenamefont
  {Gra\ss{}}, \citenamefont {Muschik}, \citenamefont {Celi}, \citenamefont
  {Chhajlany},\ and\ \citenamefont {Lewenstein}}]{grass2015}%
  \BibitemOpen
  \bibfield  {author} {\bibinfo {author} {\bibfnamefont {T.}~\bibnamefont
  {Gra\ss{}}}, \bibinfo {author} {\bibfnamefont {C.}~\bibnamefont {Muschik}},
  \bibinfo {author} {\bibfnamefont {A.}~\bibnamefont {Celi}}, \bibinfo {author}
  {\bibfnamefont {R.~W.}\ \bibnamefont {Chhajlany}}, \ and\ \bibinfo {author}
  {\bibfnamefont {M.}~\bibnamefont {Lewenstein}},\ }\href {\doibase
  10.1103/PhysRevA.91.063612} {\bibfield  {journal} {\bibinfo  {journal} {Phys.
  Rev. A}\ }\textbf {\bibinfo {volume} {91}},\ \bibinfo {pages} {063612}
  (\bibinfo {year} {2015})}\BibitemShut {NoStop}%
\bibitem [{\citenamefont {Gra\ss{}}\ \emph {et~al.}(2018)\citenamefont
  {Gra\ss{}}, \citenamefont {Celi}, \citenamefont {Pagano},\ and\ \citenamefont
  {Lewenstein}}]{grass2018}%
  \BibitemOpen
  \bibfield  {author} {\bibinfo {author} {\bibfnamefont {T.}~\bibnamefont
  {Gra\ss{}}}, \bibinfo {author} {\bibfnamefont {A.}~\bibnamefont {Celi}},
  \bibinfo {author} {\bibfnamefont {G.}~\bibnamefont {Pagano}}, \ and\ \bibinfo
  {author} {\bibfnamefont {M.}~\bibnamefont {Lewenstein}},\ }\href {\doibase
  10.1103/PhysRevA.97.010302} {\bibfield  {journal} {\bibinfo  {journal} {Phys.
  Rev. A}\ }\textbf {\bibinfo {volume} {97}},\ \bibinfo {pages} {010302}
  (\bibinfo {year} {2018})}\BibitemShut {NoStop}%
\bibitem [{\citenamefont {Hauke}\ \emph {et~al.}(2013)\citenamefont {Hauke},
  \citenamefont {Marcos}, \citenamefont {Dalmonte},\ and\ \citenamefont
  {Zoller}}]{hauke2013quantum}%
  \BibitemOpen
  \bibfield  {author} {\bibinfo {author} {\bibfnamefont {P.}~\bibnamefont
  {Hauke}}, \bibinfo {author} {\bibfnamefont {D.}~\bibnamefont {Marcos}},
  \bibinfo {author} {\bibfnamefont {M.}~\bibnamefont {Dalmonte}}, \ and\
  \bibinfo {author} {\bibfnamefont {P.}~\bibnamefont {Zoller}},\ }\href
  {\doibase 10.1103/PhysRevX.3.041018} {\bibfield  {journal} {\bibinfo
  {journal} {Phys. Rev. X}\ }\textbf {\bibinfo {volume} {3}},\ \bibinfo {pages}
  {041018} (\bibinfo {year} {2013})}\BibitemShut {NoStop}%
\bibitem [{\citenamefont {Bermudez}\ \emph {et~al.}(2011)\citenamefont
  {Bermudez}, \citenamefont {Schaetz},\ and\ \citenamefont
  {Porras}}]{bermudez2011synthetic}%
  \BibitemOpen
  \bibfield  {author} {\bibinfo {author} {\bibfnamefont {A.}~\bibnamefont
  {Bermudez}}, \bibinfo {author} {\bibfnamefont {T.}~\bibnamefont {Schaetz}}, \
  and\ \bibinfo {author} {\bibfnamefont {D.}~\bibnamefont {Porras}},\ }\href
  {\doibase 10.1103/PhysRevLett.107.150501} {\bibfield  {journal} {\bibinfo
  {journal} {Phys. Rev. Lett.}\ }\textbf {\bibinfo {volume} {107}},\ \bibinfo
  {pages} {150501} (\bibinfo {year} {2011})}\BibitemShut {NoStop}%
\bibitem [{\citenamefont {Bermudez}\ \emph {et~al.}(2012)\citenamefont
  {Bermudez}, \citenamefont {Schaetz},\ and\ \citenamefont
  {Porras}}]{bermudez2012photon}%
  \BibitemOpen
  \bibfield  {author} {\bibinfo {author} {\bibfnamefont {A.}~\bibnamefont
  {Bermudez}}, \bibinfo {author} {\bibfnamefont {T.}~\bibnamefont {Schaetz}}, \
  and\ \bibinfo {author} {\bibfnamefont {D.}~\bibnamefont {Porras}},\
  }\href@noop {} {\bibfield  {journal} {\bibinfo  {journal} {New Journal of
  Physics}\ }\textbf {\bibinfo {volume} {14}},\ \bibinfo {pages} {053049}
  (\bibinfo {year} {2012})}\BibitemShut {NoStop}%
\bibitem [{\citenamefont {Vermersch}\ \emph {et~al.}(2016)\citenamefont
  {Vermersch}, \citenamefont {Ramos}, \citenamefont {Hauke},\ and\
  \citenamefont {Zoller}}]{vermersch2016implementation}%
  \BibitemOpen
  \bibfield  {author} {\bibinfo {author} {\bibfnamefont {B.}~\bibnamefont
  {Vermersch}}, \bibinfo {author} {\bibfnamefont {T.}~\bibnamefont {Ramos}},
  \bibinfo {author} {\bibfnamefont {P.}~\bibnamefont {Hauke}}, \ and\ \bibinfo
  {author} {\bibfnamefont {P.}~\bibnamefont {Zoller}},\ }\href@noop {}
  {\bibfield  {journal} {\bibinfo  {journal} {Physical Review A}\ }\textbf
  {\bibinfo {volume} {93}},\ \bibinfo {pages} {063830} (\bibinfo {year}
  {2016})}\BibitemShut {NoStop}%
\bibitem [{\citenamefont {Aharonov}\ and\ \citenamefont
  {Bohm}(1959)}]{aharonov1959}%
  \BibitemOpen
  \bibfield  {author} {\bibinfo {author} {\bibfnamefont {Y.}~\bibnamefont
  {Aharonov}}\ and\ \bibinfo {author} {\bibfnamefont {D.}~\bibnamefont
  {Bohm}},\ }\href {\doibase 10.1103/PhysRev.115.485} {\bibfield  {journal}
  {\bibinfo  {journal} {Phys. Rev.}\ }\textbf {\bibinfo {volume} {115}},\
  \bibinfo {pages} {485} (\bibinfo {year} {1959})}\BibitemShut {NoStop}%
\bibitem [{\citenamefont {Roushan}(2017)}]{roushan2017chiral}%
  \BibitemOpen
  \bibfield  {author} {\bibinfo {author} {\bibfnamefont {N.~C. M. A. e.~a.}\
  \bibnamefont {Roushan}, \bibfnamefont {P.}},\ }\href@noop {} {\bibfield
  {journal} {\bibinfo  {journal} {Nature Phys.}\ }\textbf {\bibinfo {volume}
  {13}},\ \bibinfo {pages} {146} (\bibinfo {year} {2017})}\BibitemShut
  {NoStop}%
\bibitem [{\citenamefont {Shaniv}\ \emph {et~al.}(2018)\citenamefont {Shaniv},
  \citenamefont {Manovitz}, \citenamefont {Shapira}, \citenamefont {Akerman},\
  and\ \citenamefont {Ozeri}}]{shaniv2018toward}%
  \BibitemOpen
  \bibfield  {author} {\bibinfo {author} {\bibfnamefont {R.}~\bibnamefont
  {Shaniv}}, \bibinfo {author} {\bibfnamefont {T.}~\bibnamefont {Manovitz}},
  \bibinfo {author} {\bibfnamefont {Y.}~\bibnamefont {Shapira}}, \bibinfo
  {author} {\bibfnamefont {N.}~\bibnamefont {Akerman}}, \ and\ \bibinfo
  {author} {\bibfnamefont {R.}~\bibnamefont {Ozeri}},\ }\href@noop {}
  {\bibfield  {journal} {\bibinfo  {journal} {Physical Review Letters}\
  }\textbf {\bibinfo {volume} {120}},\ \bibinfo {pages} {243603} (\bibinfo
  {year} {2018})}\BibitemShut {NoStop}%
\bibitem [{\citenamefont {Magnus}(1954)}]{magnus1954onthe}%
  \BibitemOpen
  \bibfield  {author} {\bibinfo {author} {\bibfnamefont {W.}~\bibnamefont
  {Magnus}},\ }\href {\doibase 10.1002/cpa.3160070404} {\bibfield  {journal}
  {\bibinfo  {journal} {Communications on Pure and Applied Mathematics}\
  }\textbf {\bibinfo {volume} {7}},\ \bibinfo {pages} {649} (\bibinfo {year}
  {1954})},\ \Eprint
  {http://arxiv.org/abs/https://onlinelibrary.wiley.com/doi/pdf/10.1002/cpa.3160070404}
  {https://onlinelibrary.wiley.com/doi/pdf/10.1002/cpa.3160070404} \BibitemShut
  {NoStop}%
\bibitem [{\citenamefont {Blanes}\ \emph {et~al.}(2010)\citenamefont {Blanes},
  \citenamefont {Casas}, \citenamefont {Oteo},\ and\ \citenamefont
  {Ros}}]{blanes2010pedagogical}%
  \BibitemOpen
  \bibfield  {author} {\bibinfo {author} {\bibfnamefont {S.}~\bibnamefont
  {Blanes}}, \bibinfo {author} {\bibfnamefont {F.}~\bibnamefont {Casas}},
  \bibinfo {author} {\bibfnamefont {J.~A.}\ \bibnamefont {Oteo}}, \ and\
  \bibinfo {author} {\bibfnamefont {J.}~\bibnamefont {Ros}},\ }\href {\doibase
  10.1088/0143-0807/31/4/020} {\bibfield  {journal} {\bibinfo  {journal}
  {European Journal of Physics}\ }\textbf {\bibinfo {volume} {31}},\ \bibinfo
  {pages} {907} (\bibinfo {year} {2010})}\BibitemShut {NoStop}%
\bibitem [{\citenamefont {Shankar}(2017)}]{shankar2017exact}%
  \BibitemOpen
  \bibfield  {author} {\bibinfo {author} {\bibfnamefont {R.}~\bibnamefont
  {Shankar}},\ }\enquote {\bibinfo {title} {Exact solution of the
  two–dimensional ising model},}\ in\ \href {\doibase
  10.1017/9781139044349.009} {\emph {\bibinfo {booktitle} {Quantum Field Theory
  and Condensed Matter: An Introduction}}}\ (\bibinfo  {publisher} {Cambridge
  University Press},\ \bibinfo {year} {2017})\ p.\ \bibinfo {pages}
  {114–142}\BibitemShut {NoStop}%
\bibitem [{\citenamefont {Viefers}\ \emph {et~al.}(2004)\citenamefont
  {Viefers}, \citenamefont {Koskinen}, \citenamefont {Deo},\ and\ \citenamefont
  {Manninen}}]{viefers2004quantum}%
  \BibitemOpen
  \bibfield  {author} {\bibinfo {author} {\bibfnamefont {S.}~\bibnamefont
  {Viefers}}, \bibinfo {author} {\bibfnamefont {P.}~\bibnamefont {Koskinen}},
  \bibinfo {author} {\bibfnamefont {P.~S.}\ \bibnamefont {Deo}}, \ and\
  \bibinfo {author} {\bibfnamefont {M.}~\bibnamefont {Manninen}},\ }\href@noop
  {} {\bibfield  {journal} {\bibinfo  {journal} {Physica E: Low-dimensional
  Systems and Nanostructures}\ }\textbf {\bibinfo {volume} {21}},\ \bibinfo
  {pages} {1} (\bibinfo {year} {2004})}\BibitemShut {NoStop}%
\bibitem [{\citenamefont {M.~Büttiker}\ and\ \citenamefont
  {Landauer}(1983)}]{butiker1983josephson}%
  \BibitemOpen
  \bibfield  {author} {\bibinfo {author} {\bibfnamefont {Y.~I.}\ \bibnamefont
  {M.~Büttiker}}\ and\ \bibinfo {author} {\bibfnamefont {R.}~\bibnamefont
  {Landauer}},\ }\href@noop {} {\bibfield  {journal} {\bibinfo  {journal}
  {Physics Letters A}\ }\textbf {\bibinfo {volume} {96}},\ \bibinfo {pages}
  {365} (\bibinfo {year} {1983})}\BibitemShut {NoStop}%
\bibitem [{\citenamefont {H.~F.~Cheung}(1988)}]{hofai1988persistent}%
  \BibitemOpen
  \bibfield  {author} {\bibinfo {author} {\bibfnamefont {E.~K. R. W. H.~S.}\
  \bibnamefont {H.~F.~Cheung}, \bibfnamefont {Y.~Gefen}},\ }\href@noop {}
  {\bibfield  {journal} {\bibinfo  {journal} {Phys. Rev. A}\ }\textbf {\bibinfo
  {volume} {37}},\ \bibinfo {pages} {6050} (\bibinfo {year}
  {1988})}\BibitemShut {NoStop}%
\bibitem [{\citenamefont {Ant{\'{o}}nio}\ \emph {et~al.}(2013)\citenamefont
  {Ant{\'{o}}nio}, \citenamefont {Lopes},\ and\ \citenamefont
  {Dias}}]{antonio2013transport}%
  \BibitemOpen
  \bibfield  {author} {\bibinfo {author} {\bibfnamefont {B.~A.~Z.}\
  \bibnamefont {Ant{\'{o}}nio}}, \bibinfo {author} {\bibfnamefont {A.~A.}\
  \bibnamefont {Lopes}}, \ and\ \bibinfo {author} {\bibfnamefont {R.~G.}\
  \bibnamefont {Dias}},\ }\href {\doibase 10.1088/0143-0807/34/4/831}
  {\bibfield  {journal} {\bibinfo  {journal} {European Journal of Physics}\
  }\textbf {\bibinfo {volume} {34}},\ \bibinfo {pages} {831} (\bibinfo {year}
  {2013})}\BibitemShut {NoStop}%
\bibitem [{\citenamefont {Bloch}(1929)}]{bloch1929quantenmechanik}%
  \BibitemOpen
  \bibfield  {author} {\bibinfo {author} {\bibfnamefont {F.}~\bibnamefont
  {Bloch}},\ }\href@noop {} {\bibfield  {journal} {\bibinfo  {journal}
  {Zeitschrift f{\"u}r physik}\ }\textbf {\bibinfo {volume} {52}},\ \bibinfo
  {pages} {555} (\bibinfo {year} {1929})}\BibitemShut {NoStop}%
\bibitem [{\citenamefont {Amico}\ \emph {et~al.}(2008)\citenamefont {Amico},
  \citenamefont {Fazio}, \citenamefont {Osterloh},\ and\ \citenamefont
  {Vedral}}]{amico2008entanglement}%
  \BibitemOpen
  \bibfield  {author} {\bibinfo {author} {\bibfnamefont {L.}~\bibnamefont
  {Amico}}, \bibinfo {author} {\bibfnamefont {R.}~\bibnamefont {Fazio}},
  \bibinfo {author} {\bibfnamefont {A.}~\bibnamefont {Osterloh}}, \ and\
  \bibinfo {author} {\bibfnamefont {V.}~\bibnamefont {Vedral}},\ }\href@noop {}
  {\bibfield  {journal} {\bibinfo  {journal} {Reviews of modern physics}\
  }\textbf {\bibinfo {volume} {80}},\ \bibinfo {pages} {517} (\bibinfo {year}
  {2008})}\BibitemShut {NoStop}%
\bibitem [{\citenamefont {Diep}\ \emph {et~al.}(2013)\citenamefont {Diep} \emph
  {et~al.}}]{diep2013frustrated}%
  \BibitemOpen
  \bibfield  {author} {\bibinfo {author} {\bibfnamefont {H.}~\bibnamefont
  {Diep}} \emph {et~al.},\ }\href@noop {} {\emph {\bibinfo {title} {Frustrated
  spin systems}}}\ (\bibinfo  {publisher} {World Scientific},\ \bibinfo {year}
  {2013})\BibitemShut {NoStop}%
\bibitem [{\citenamefont {Majumdar}\ and\ \citenamefont
  {Ghosh}(1969)}]{majumdar1969next}%
  \BibitemOpen
  \bibfield  {author} {\bibinfo {author} {\bibfnamefont {C.~K.}\ \bibnamefont
  {Majumdar}}\ and\ \bibinfo {author} {\bibfnamefont {D.~K.}\ \bibnamefont
  {Ghosh}},\ }\href@noop {} {\bibfield  {journal} {\bibinfo  {journal} {Journal
  of Mathematical Physics}\ }\textbf {\bibinfo {volume} {10}},\ \bibinfo
  {pages} {1388} (\bibinfo {year} {1969})}\BibitemShut {NoStop}%
\bibitem [{\citenamefont {Furukawa}\ \emph {et~al.}(2012)\citenamefont
  {Furukawa}, \citenamefont {Sato}, \citenamefont {Onoda},\ and\ \citenamefont
  {Furusaki}}]{furukawa2012ground}%
  \BibitemOpen
  \bibfield  {author} {\bibinfo {author} {\bibfnamefont {S.}~\bibnamefont
  {Furukawa}}, \bibinfo {author} {\bibfnamefont {M.}~\bibnamefont {Sato}},
  \bibinfo {author} {\bibfnamefont {S.}~\bibnamefont {Onoda}}, \ and\ \bibinfo
  {author} {\bibfnamefont {A.}~\bibnamefont {Furusaki}},\ }\href@noop {}
  {\bibfield  {journal} {\bibinfo  {journal} {Physical Review B}\ }\textbf
  {\bibinfo {volume} {86}},\ \bibinfo {pages} {094417} (\bibinfo {year}
  {2012})}\BibitemShut {NoStop}%
\bibitem [{\citenamefont {Furukawa}\ \emph {et~al.}(2010)\citenamefont
  {Furukawa}, \citenamefont {Sato},\ and\ \citenamefont
  {Onoda}}]{furukawa2010chiral}%
  \BibitemOpen
  \bibfield  {author} {\bibinfo {author} {\bibfnamefont {S.}~\bibnamefont
  {Furukawa}}, \bibinfo {author} {\bibfnamefont {M.}~\bibnamefont {Sato}}, \
  and\ \bibinfo {author} {\bibfnamefont {S.}~\bibnamefont {Onoda}},\
  }\href@noop {} {\bibfield  {journal} {\bibinfo  {journal} {Physical Review
  Letters}\ }\textbf {\bibinfo {volume} {105}},\ \bibinfo {pages} {257205}
  (\bibinfo {year} {2010})}\BibitemShut {NoStop}%
\bibitem [{\citenamefont {Hikihara}\ \emph {et~al.}(2008)\citenamefont
  {Hikihara}, \citenamefont {Kecke}, \citenamefont {Momoi},\ and\ \citenamefont
  {Furusaki}}]{hikihara2008vector}%
  \BibitemOpen
  \bibfield  {author} {\bibinfo {author} {\bibfnamefont {T.}~\bibnamefont
  {Hikihara}}, \bibinfo {author} {\bibfnamefont {L.}~\bibnamefont {Kecke}},
  \bibinfo {author} {\bibfnamefont {T.}~\bibnamefont {Momoi}}, \ and\ \bibinfo
  {author} {\bibfnamefont {A.}~\bibnamefont {Furusaki}},\ }\href@noop {}
  {\bibfield  {journal} {\bibinfo  {journal} {Physical Review B}\ }\textbf
  {\bibinfo {volume} {78}},\ \bibinfo {pages} {144404} (\bibinfo {year}
  {2008})}\BibitemShut {NoStop}%
\bibitem [{\citenamefont {Nersesyan}\ \emph {et~al.}(1998)\citenamefont
  {Nersesyan}, \citenamefont {Gogolin},\ and\ \citenamefont
  {E{\ss}ler}}]{nersesyan1998incommensurate}%
  \BibitemOpen
  \bibfield  {author} {\bibinfo {author} {\bibfnamefont {A.~A.}\ \bibnamefont
  {Nersesyan}}, \bibinfo {author} {\bibfnamefont {A.~O.}\ \bibnamefont
  {Gogolin}}, \ and\ \bibinfo {author} {\bibfnamefont {F.~H.}\ \bibnamefont
  {E{\ss}ler}},\ }\href@noop {} {\bibfield  {journal} {\bibinfo  {journal}
  {Physical Review Letters}\ }\textbf {\bibinfo {volume} {81}},\ \bibinfo
  {pages} {910} (\bibinfo {year} {1998})}\BibitemShut {NoStop}%
\bibitem [{\citenamefont {Islam}\ \emph
  {et~al.}(2011{\natexlab{b}})\citenamefont {Islam}, \citenamefont {Edwards},
  \citenamefont {Kim}, \citenamefont {Korenblit}, \citenamefont {Noh},
  \citenamefont {Carmichael}, \citenamefont {Lin}, \citenamefont {Duan},
  \citenamefont {Wang}, \citenamefont {Freericks} \emph
  {et~al.}}]{islam2011onset}%
  \BibitemOpen
  \bibfield  {author} {\bibinfo {author} {\bibfnamefont {R.}~\bibnamefont
  {Islam}}, \bibinfo {author} {\bibfnamefont {E.}~\bibnamefont {Edwards}},
  \bibinfo {author} {\bibfnamefont {K.}~\bibnamefont {Kim}}, \bibinfo {author}
  {\bibfnamefont {S.}~\bibnamefont {Korenblit}}, \bibinfo {author}
  {\bibfnamefont {C.}~\bibnamefont {Noh}}, \bibinfo {author} {\bibfnamefont
  {H.}~\bibnamefont {Carmichael}}, \bibinfo {author} {\bibfnamefont {G.-D.}\
  \bibnamefont {Lin}}, \bibinfo {author} {\bibfnamefont {L.-M.}\ \bibnamefont
  {Duan}}, \bibinfo {author} {\bibfnamefont {C.-C.~J.}\ \bibnamefont {Wang}},
  \bibinfo {author} {\bibfnamefont {J.}~\bibnamefont {Freericks}},  \emph
  {et~al.},\ }\href@noop {} {\bibfield  {journal} {\bibinfo  {journal} {Nature
  communications}\ }\textbf {\bibinfo {volume} {2}},\ \bibinfo {pages} {1}
  (\bibinfo {year} {2011}{\natexlab{b}})}\BibitemShut {NoStop}%
\bibitem [{\citenamefont {Brydges}\ \emph {et~al.}(2019)\citenamefont
  {Brydges}, \citenamefont {Elben}, \citenamefont {Jurcevic}, \citenamefont
  {Vermersch}, \citenamefont {Maier}, \citenamefont {Lanyon}, \citenamefont
  {Zoller}, \citenamefont {Blatt},\ and\ \citenamefont {Roos}}]{brydges2019}%
  \BibitemOpen
  \bibfield  {author} {\bibinfo {author} {\bibfnamefont {T.}~\bibnamefont
  {Brydges}}, \bibinfo {author} {\bibfnamefont {A.}~\bibnamefont {Elben}},
  \bibinfo {author} {\bibfnamefont {P.}~\bibnamefont {Jurcevic}}, \bibinfo
  {author} {\bibfnamefont {B.}~\bibnamefont {Vermersch}}, \bibinfo {author}
  {\bibfnamefont {C.}~\bibnamefont {Maier}}, \bibinfo {author} {\bibfnamefont
  {B.~P.}\ \bibnamefont {Lanyon}}, \bibinfo {author} {\bibfnamefont
  {P.}~\bibnamefont {Zoller}}, \bibinfo {author} {\bibfnamefont
  {R.}~\bibnamefont {Blatt}}, \ and\ \bibinfo {author} {\bibfnamefont {C.~F.}\
  \bibnamefont {Roos}},\ }\href {\doibase 10.1126/science.aau4963} {\bibfield
  {journal} {\bibinfo  {journal} {Science}\ }\textbf {\bibinfo {volume}
  {364}},\ \bibinfo {pages} {260} (\bibinfo {year} {2019})},\ \Eprint
  {http://arxiv.org/abs/https://science.sciencemag.org/content/364/6437/260.full.pdf}
  {https://science.sciencemag.org/content/364/6437/260.full.pdf} \BibitemShut
  {NoStop}%
\bibitem [{\citenamefont {Anisimovas}\ \emph {et~al.}(2016)\citenamefont
  {Anisimovas}, \citenamefont {Ra{\v{c}}i{\=u}nas}, \citenamefont
  {Str{\"a}ter}, \citenamefont {Eckardt}, \citenamefont {Spielman},\ and\
  \citenamefont {Juzeli{\=u}nas}}]{anisimovas2016semisynthetic}%
  \BibitemOpen
  \bibfield  {author} {\bibinfo {author} {\bibfnamefont {E.}~\bibnamefont
  {Anisimovas}}, \bibinfo {author} {\bibfnamefont {M.}~\bibnamefont
  {Ra{\v{c}}i{\=u}nas}}, \bibinfo {author} {\bibfnamefont {C.}~\bibnamefont
  {Str{\"a}ter}}, \bibinfo {author} {\bibfnamefont {A.}~\bibnamefont
  {Eckardt}}, \bibinfo {author} {\bibfnamefont {I.}~\bibnamefont {Spielman}}, \
  and\ \bibinfo {author} {\bibfnamefont {G.}~\bibnamefont {Juzeli{\=u}nas}},\
  }\href@noop {} {\bibfield  {journal} {\bibinfo  {journal} {Physical Review
  A}\ }\textbf {\bibinfo {volume} {94}},\ \bibinfo {pages} {063632} (\bibinfo
  {year} {2016})}\BibitemShut {NoStop}%
\bibitem [{\citenamefont {Suszalski}\ and\ \citenamefont
  {Zakrzewski}(2016)}]{suszalski2016different}%
  \BibitemOpen
  \bibfield  {author} {\bibinfo {author} {\bibfnamefont {D.}~\bibnamefont
  {Suszalski}}\ and\ \bibinfo {author} {\bibfnamefont {J.}~\bibnamefont
  {Zakrzewski}},\ }\href@noop {} {\bibfield  {journal} {\bibinfo  {journal}
  {Physical Review A}\ }\textbf {\bibinfo {volume} {94}},\ \bibinfo {pages}
  {033602} (\bibinfo {year} {2016})}\BibitemShut {NoStop}%
\bibitem [{\citenamefont {An}\ \emph {et~al.}(2018)\citenamefont {An},
  \citenamefont {Meier},\ and\ \citenamefont {Gadway}}]{an2018engineering}%
  \BibitemOpen
  \bibfield  {author} {\bibinfo {author} {\bibfnamefont {F.~A.}\ \bibnamefont
  {An}}, \bibinfo {author} {\bibfnamefont {E.~J.}\ \bibnamefont {Meier}}, \
  and\ \bibinfo {author} {\bibfnamefont {B.}~\bibnamefont {Gadway}},\
  }\href@noop {} {\bibfield  {journal} {\bibinfo  {journal} {Physical Review
  X}\ }\textbf {\bibinfo {volume} {8}},\ \bibinfo {pages} {031045} (\bibinfo
  {year} {2018})}\BibitemShut {NoStop}%
\bibitem [{\citenamefont {Cabedo}\ \emph {et~al.}(2020)\citenamefont {Cabedo},
  \citenamefont {Claramunt}, \citenamefont {Mompart}, \citenamefont
  {Ahufinger},\ and\ \citenamefont {Celi}}]{cabedo2020effective}%
  \BibitemOpen
  \bibfield  {author} {\bibinfo {author} {\bibfnamefont {J.}~\bibnamefont
  {Cabedo}}, \bibinfo {author} {\bibfnamefont {J.}~\bibnamefont {Claramunt}},
  \bibinfo {author} {\bibfnamefont {J.}~\bibnamefont {Mompart}}, \bibinfo
  {author} {\bibfnamefont {V.}~\bibnamefont {Ahufinger}}, \ and\ \bibinfo
  {author} {\bibfnamefont {A.}~\bibnamefont {Celi}},\ }\href@noop {} {\bibfield
   {journal} {\bibinfo  {journal} {arXiv preprint arXiv:2003.04154}\ }
  (\bibinfo {year} {2020})}\BibitemShut {NoStop}%
\bibitem [{\citenamefont {N.~Zhao}\ and\ \citenamefont
  {Sun}(2009)}]{zhao2009mobius1}%
  \BibitemOpen
  \bibfield  {author} {\bibinfo {author} {\bibfnamefont {S.~Y.}\ \bibnamefont
  {N.~Zhao}, \bibfnamefont {H.~Dong}}\ and\ \bibinfo {author} {\bibfnamefont
  {C.~P.}\ \bibnamefont {Sun}},\ }\href@noop {} {\bibfield  {journal} {\bibinfo
   {journal} {Phys. Rev. B}\ }\textbf {\bibinfo {volume} {79}},\ \bibinfo
  {pages} {125440} (\bibinfo {year} {2009})}\BibitemShut {NoStop}%
\bibitem [{\citenamefont {Z.~L.~Guo}\ and\ \citenamefont
  {Sun}(2009)}]{guo2009mobius2}%
  \BibitemOpen
  \bibfield  {author} {\bibinfo {author} {\bibfnamefont {H.~D.}\ \bibnamefont
  {Z.~L.~Guo}, \bibfnamefont {Z.~R.~Gong}}\ and\ \bibinfo {author}
  {\bibfnamefont {C.~P.}\ \bibnamefont {Sun}},\ }\href@noop {} {\bibfield
  {journal} {\bibinfo  {journal} {Phys. Rev. B}\ }\textbf {\bibinfo {volume}
  {80}},\ \bibinfo {pages} {195310} (\bibinfo {year} {2009})}\BibitemShut
  {NoStop}%
\bibitem [{\citenamefont {J.~Ningyuan}\ and\ \citenamefont
  {Simon}(2015)}]{ningyuan2015mobius3}%
  \BibitemOpen
  \bibfield  {author} {\bibinfo {author} {\bibfnamefont {A.~S. D.~S.}\
  \bibnamefont {J.~Ningyuan}, \bibfnamefont {C.~Owens}}\ and\ \bibinfo {author}
  {\bibfnamefont {J.}~\bibnamefont {Simon}},\ }\href@noop {} {\bibfield
  {journal} {\bibinfo  {journal} {Phys. Rev. X}\ }\textbf {\bibinfo {volume}
  {5}},\ \bibinfo {pages} {021031} (\bibinfo {year} {2015})}\BibitemShut
  {NoStop}%
\bibitem [{\citenamefont {Yakubo}\ \emph {et~al.}(2003)\citenamefont {Yakubo},
  \citenamefont {Avishai},\ and\ \citenamefont {Cohen}}]{yakubo2003}%
  \BibitemOpen
  \bibfield  {author} {\bibinfo {author} {\bibfnamefont {K.}~\bibnamefont
  {Yakubo}}, \bibinfo {author} {\bibfnamefont {Y.}~\bibnamefont {Avishai}}, \
  and\ \bibinfo {author} {\bibfnamefont {D.}~\bibnamefont {Cohen}},\ }\href
  {\doibase 10.1103/PhysRevB.67.125319} {\bibfield  {journal} {\bibinfo
  {journal} {Phys. Rev. B}\ }\textbf {\bibinfo {volume} {67}},\ \bibinfo
  {pages} {125319} (\bibinfo {year} {2003})}\BibitemShut {NoStop}%
\bibitem [{\citenamefont {Bergholtz}\ and\ \citenamefont
  {Karlhede}(2005)}]{bergholtz2005half}%
  \BibitemOpen
  \bibfield  {author} {\bibinfo {author} {\bibfnamefont {E.~J.}\ \bibnamefont
  {Bergholtz}}\ and\ \bibinfo {author} {\bibfnamefont {A.}~\bibnamefont
  {Karlhede}},\ }\href {\doibase 10.1103/PhysRevLett.94.026802} {\bibfield
  {journal} {\bibinfo  {journal} {Phys. Rev. Lett.}\ }\textbf {\bibinfo
  {volume} {94}},\ \bibinfo {pages} {026802} (\bibinfo {year}
  {2005})}\BibitemShut {NoStop}%
\bibitem [{\citenamefont {Bernevig}\ and\ \citenamefont
  {Regnault}(2012)}]{bernevig2012thintorus}%
  \BibitemOpen
  \bibfield  {author} {\bibinfo {author} {\bibfnamefont {B.~A.}\ \bibnamefont
  {Bernevig}}\ and\ \bibinfo {author} {\bibfnamefont {N.}~\bibnamefont
  {Regnault}},\ }\href@noop {} {\  (\bibinfo {year} {2012})},\ \Eprint
  {http://arxiv.org/abs/1204.5682} {arXiv:1204.5682 [cond-mat.str-el]}
  \BibitemShut {NoStop}%
\bibitem [{\citenamefont {Boada}\ \emph {et~al.}(2012)\citenamefont {Boada},
  \citenamefont {Celi}, \citenamefont {Latorre},\ and\ \citenamefont
  {Lewenstein}}]{Boada2012}%
  \BibitemOpen
  \bibfield  {author} {\bibinfo {author} {\bibfnamefont {O.}~\bibnamefont
  {Boada}}, \bibinfo {author} {\bibfnamefont {A.}~\bibnamefont {Celi}},
  \bibinfo {author} {\bibfnamefont {J.~I.}\ \bibnamefont {Latorre}}, \ and\
  \bibinfo {author} {\bibfnamefont {M.}~\bibnamefont {Lewenstein}},\ }\href
  {\doibase 10.1103/PhysRevLett.108.133001} {\bibfield  {journal} {\bibinfo
  {journal} {Phys. Rev. Lett.}\ }\textbf {\bibinfo {volume} {108}},\ \bibinfo
  {pages} {133001} (\bibinfo {year} {2012})}\BibitemShut {NoStop}%
\bibitem [{\citenamefont {Livi}\ \emph {et~al.}(2016)\citenamefont {Livi},
  \citenamefont {Cappellini}, \citenamefont {Diem}, \citenamefont {Franchi},
  \citenamefont {Clivati}, \citenamefont {Frittelli}, \citenamefont {Levi},
  \citenamefont {Calonico}, \citenamefont {Catani}, \citenamefont {Inguscio},\
  and\ \citenamefont {Fallani}}]{celi2016}%
  \BibitemOpen
  \bibfield  {author} {\bibinfo {author} {\bibfnamefont {L.~F.}\ \bibnamefont
  {Livi}}, \bibinfo {author} {\bibfnamefont {G.}~\bibnamefont {Cappellini}},
  \bibinfo {author} {\bibfnamefont {M.}~\bibnamefont {Diem}}, \bibinfo {author}
  {\bibfnamefont {L.}~\bibnamefont {Franchi}}, \bibinfo {author} {\bibfnamefont
  {C.}~\bibnamefont {Clivati}}, \bibinfo {author} {\bibfnamefont
  {M.}~\bibnamefont {Frittelli}}, \bibinfo {author} {\bibfnamefont
  {F.}~\bibnamefont {Levi}}, \bibinfo {author} {\bibfnamefont {D.}~\bibnamefont
  {Calonico}}, \bibinfo {author} {\bibfnamefont {J.}~\bibnamefont {Catani}},
  \bibinfo {author} {\bibfnamefont {M.}~\bibnamefont {Inguscio}}, \ and\
  \bibinfo {author} {\bibfnamefont {L.}~\bibnamefont {Fallani}},\ }\href
  {\doibase 10.1103/PhysRevLett.117.220401} {\bibfield  {journal} {\bibinfo
  {journal} {Phys. Rev. Lett.}\ }\textbf {\bibinfo {volume} {117}},\ \bibinfo
  {pages} {220401} (\bibinfo {year} {2016})}\BibitemShut {NoStop}%
\bibitem [{\citenamefont {Anderson}(1958)}]{anderson1958localization}%
  \BibitemOpen
  \bibfield  {author} {\bibinfo {author} {\bibfnamefont {P.~W.}\ \bibnamefont
  {Anderson}},\ }\href@noop {} {\bibfield  {journal} {\bibinfo  {journal}
  {Phys. Rev.}\ }\textbf {\bibinfo {volume} {109}},\ \bibinfo {pages} {1492}
  (\bibinfo {year} {1958})}\BibitemShut {NoStop}%
\bibitem [{\citenamefont {James}(2000)}]{james2000quantum}%
  \BibitemOpen
  \bibfield  {author} {\bibinfo {author} {\bibfnamefont {D.~F.}\ \bibnamefont
  {James}},\ }\href@noop {} {\bibfield  {journal} {\bibinfo  {journal} {Quantum
  Computation and Quantum Information Theory: Reprint Volume with Introductory
  Notes for ISI TMR Network School, 12-23 July 1999, Villa Gualino, Torino,
  Italy}\ }\textbf {\bibinfo {volume} {66}},\ \bibinfo {pages} {345} (\bibinfo
  {year} {2000})}\BibitemShut {NoStop}%
\bibitem [{\citenamefont {Pagano}\ \emph {et~al.}(2018)\citenamefont {Pagano},
  \citenamefont {Hess}, \citenamefont {Kaplan}, \citenamefont {Tan},
  \citenamefont {Richerme}, \citenamefont {Becker}, \citenamefont
  {Kyprianidis}, \citenamefont {Zhang}, \citenamefont {Birckelbaw},
  \citenamefont {Hernandez} \emph {et~al.}}]{pagano2018cryogenic}%
  \BibitemOpen
  \bibfield  {author} {\bibinfo {author} {\bibfnamefont {G.}~\bibnamefont
  {Pagano}}, \bibinfo {author} {\bibfnamefont {P.}~\bibnamefont {Hess}},
  \bibinfo {author} {\bibfnamefont {H.}~\bibnamefont {Kaplan}}, \bibinfo
  {author} {\bibfnamefont {W.}~\bibnamefont {Tan}}, \bibinfo {author}
  {\bibfnamefont {P.}~\bibnamefont {Richerme}}, \bibinfo {author}
  {\bibfnamefont {P.}~\bibnamefont {Becker}}, \bibinfo {author} {\bibfnamefont
  {A.}~\bibnamefont {Kyprianidis}}, \bibinfo {author} {\bibfnamefont
  {J.}~\bibnamefont {Zhang}}, \bibinfo {author} {\bibfnamefont
  {E.}~\bibnamefont {Birckelbaw}}, \bibinfo {author} {\bibfnamefont
  {M.}~\bibnamefont {Hernandez}},  \emph {et~al.},\ }\href@noop {} {\bibfield
  {journal} {\bibinfo  {journal} {Quantum Science and Technology}\ }\textbf
  {\bibinfo {volume} {4}},\ \bibinfo {pages} {014004} (\bibinfo {year}
  {2018})}\BibitemShut {NoStop}%
\bibitem [{\citenamefont {Lin}\ \emph {et~al.}(2009{\natexlab{b}})\citenamefont
  {Lin}, \citenamefont {Zhu}, \citenamefont {Islam}, \citenamefont {Kim},
  \citenamefont {Chang}, \citenamefont {Korenblit}, \citenamefont {Monroe},\
  and\ \citenamefont {Duan}}]{lin2009}%
  \BibitemOpen
  \bibfield  {author} {\bibinfo {author} {\bibfnamefont {G.-D.}\ \bibnamefont
  {Lin}}, \bibinfo {author} {\bibfnamefont {S.-L.}\ \bibnamefont {Zhu}},
  \bibinfo {author} {\bibfnamefont {R.}~\bibnamefont {Islam}}, \bibinfo
  {author} {\bibfnamefont {K.}~\bibnamefont {Kim}}, \bibinfo {author}
  {\bibfnamefont {M.-S.}\ \bibnamefont {Chang}}, \bibinfo {author}
  {\bibfnamefont {S.}~\bibnamefont {Korenblit}}, \bibinfo {author}
  {\bibfnamefont {C.}~\bibnamefont {Monroe}}, \ and\ \bibinfo {author}
  {\bibfnamefont {L.-M.}\ \bibnamefont {Duan}},\ }\href {\doibase
  10.1209/0295-5075/86/60004} {\bibfield  {journal} {\bibinfo  {journal} {{EPL}
  (Europhysics Letters)}\ }\textbf {\bibinfo {volume} {86}},\ \bibinfo {pages}
  {60004} (\bibinfo {year} {2009}{\natexlab{b}})}\BibitemShut {NoStop}%
\bibitem [{\citenamefont {Doret}\ \emph {et~al.}(2012)\citenamefont {Doret},
  \citenamefont {Amini}, \citenamefont {Wright}, \citenamefont {Volin},
  \citenamefont {Killian}, \citenamefont {Ozakin}, \citenamefont {Denison},
  \citenamefont {Hayden}, \citenamefont {Pai}, \citenamefont {Slusher},\ and\
  \citenamefont {Harter}}]{doret2012}%
  \BibitemOpen
  \bibfield  {author} {\bibinfo {author} {\bibfnamefont {S.~C.}\ \bibnamefont
  {Doret}}, \bibinfo {author} {\bibfnamefont {J.~M.}\ \bibnamefont {Amini}},
  \bibinfo {author} {\bibfnamefont {K.}~\bibnamefont {Wright}}, \bibinfo
  {author} {\bibfnamefont {C.}~\bibnamefont {Volin}}, \bibinfo {author}
  {\bibfnamefont {T.}~\bibnamefont {Killian}}, \bibinfo {author} {\bibfnamefont
  {A.}~\bibnamefont {Ozakin}}, \bibinfo {author} {\bibfnamefont
  {D.}~\bibnamefont {Denison}}, \bibinfo {author} {\bibfnamefont
  {H.}~\bibnamefont {Hayden}}, \bibinfo {author} {\bibfnamefont {C.-S.}\
  \bibnamefont {Pai}}, \bibinfo {author} {\bibfnamefont {R.~E.}\ \bibnamefont
  {Slusher}}, \ and\ \bibinfo {author} {\bibfnamefont {A.~W.}\ \bibnamefont
  {Harter}},\ }\href {\doibase 10.1088/1367-2630/14/7/073012} {\bibfield
  {journal} {\bibinfo  {journal} {New Journal of Physics}\ }\textbf {\bibinfo
  {volume} {14}},\ \bibinfo {pages} {073012} (\bibinfo {year}
  {2012})}\BibitemShut {NoStop}%
\bibitem [{\citenamefont {Friis}\ \emph {et~al.}(2018)\citenamefont {Friis},
  \citenamefont {Marty}, \citenamefont {Maier}, \citenamefont {Hempel},
  \citenamefont {Holz{\"a}pfel}, \citenamefont {Jurcevic}, \citenamefont
  {Plenio}, \citenamefont {Huber}, \citenamefont {Roos}, \citenamefont {Blatt}
  \emph {et~al.}}]{friis2018observation}%
  \BibitemOpen
  \bibfield  {author} {\bibinfo {author} {\bibfnamefont {N.}~\bibnamefont
  {Friis}}, \bibinfo {author} {\bibfnamefont {O.}~\bibnamefont {Marty}},
  \bibinfo {author} {\bibfnamefont {C.}~\bibnamefont {Maier}}, \bibinfo
  {author} {\bibfnamefont {C.}~\bibnamefont {Hempel}}, \bibinfo {author}
  {\bibfnamefont {M.}~\bibnamefont {Holz{\"a}pfel}}, \bibinfo {author}
  {\bibfnamefont {P.}~\bibnamefont {Jurcevic}}, \bibinfo {author}
  {\bibfnamefont {M.~B.}\ \bibnamefont {Plenio}}, \bibinfo {author}
  {\bibfnamefont {M.}~\bibnamefont {Huber}}, \bibinfo {author} {\bibfnamefont
  {C.}~\bibnamefont {Roos}}, \bibinfo {author} {\bibfnamefont {R.}~\bibnamefont
  {Blatt}},  \emph {et~al.},\ }\href@noop {} {\bibfield  {journal} {\bibinfo
  {journal} {Physical Review X}\ }\textbf {\bibinfo {volume} {8}},\ \bibinfo
  {pages} {021012} (\bibinfo {year} {2018})}\BibitemShut {NoStop}%
\end{thebibliography}%

\section{Appendix A - Resource-efficient implementation}

In the given formulation, each $H_n$ in Eq. (\ref{genform}) requires a bichromatic field with frequencies $f_{n,\pm}=\omega_{eg}+\nu+\xi_n\pm n\Delta/2$, where care has to be taken to avoid cross-term resonances. However, for a subset of target Hamiltonians, it is possible to use less frequencies. For a pair with frequencies $f_+-f_-=n_1\Delta$ generating $H_{n_1}$, the inclusion of a single extra frequency $f_3$ such that $f_3-f_+=n_2\Delta$ would generate another two resonant coupling terms: $H_{n_2}$ and $H_{n_3}=H_{n_2+n_1}$. Given the according field amplitudes $\Omega_\pm,\Omega_3$, the effective couplings are proportional to the appropriate amplitude products: $(\Omega_{n_1})^2\sim\Omega_+\Omega_-$, $(\Omega_{n_2})^2\sim\Omega_+\Omega_3$, and $(\Omega_{n_3})^2\sim\Omega_3\Omega_-$.
We may choose $n_2=n_1$; in such a case, two terms are generated: $n_1$ and $n_2^\prime=2n_1$.

\begin{figure}
    \includegraphics[width=\columnwidth]{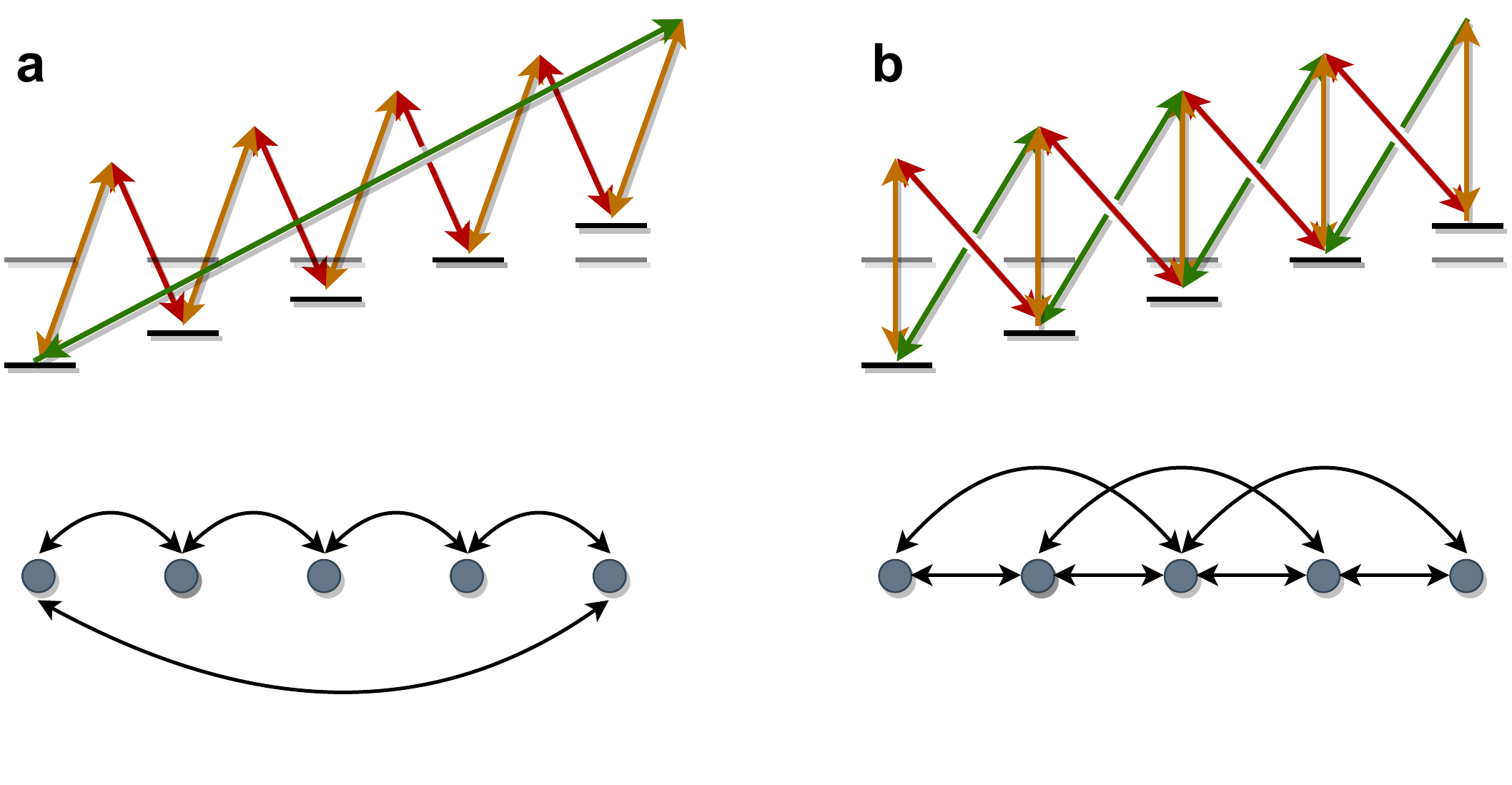}
	\caption{Resource efficient implementation. In some cases the model Hamiltonian can be implemented with less frequencies than prescribed by the general formula. Here the Aharonov-Bohm ring (a) and the triangular spin ladder (b), both analyzed above, are generated using three rather than four driving frequencies. Any frequency difference in the polychromatic field that corresponds to a frequency difference between ions in the chain will generate another coupling term.
	}
	\label{efficient}
\end{figure}

This method is naturally applicable in some cases. For instance, the triangular ladder Hamiltonian can be applied by choosing $n_2=n_1=1$.  The Aharonov-Bohm ring can be applied by choosing $n_1=1,n_2=N-1$ making use of the fact that there cannot be an $n=N$ term. We can combine the two by choosing $n_1=n_2=1$ and adding a fourth frequency $f_4=f_++(N-2)\Delta$, effectively giving the couplings: $n_1=1$, $n_2^\prime=2$, $n_3=N-2$, $n_4=N-1$; this would generate the triangular ladder with closed boundary conditions, i.e. a closed triangular band.

\section{Appendix B - Derivation of effective Hamiltonian using radial modes}

Here we derive the effective hopping Hamiltonian in presence of the multitude of modes which are necessarily in play when using the radial motional modes as interaction mediators . We assume that the laser drive generates non-negligible coupling to a multitude of radial normal-modes of motion. As in our derivations in the main text we first focus on the blue sideband driving pair. The interaction Hamiltonian in Eq. \eqref{VI} is amended to:
\begin{equation}
\begin{split}
    V_I &= i\hbar\Omega_{b}\cos\left(\frac{n\Delta}{2}t+\frac{\phi}{2}\right)\\
    &\cdot\sum_{j}a_{j}^{\dag}\sum_{k}\eta_{j,k}e^{-i\left(\omega_b-\nu_{j}\right)t}\sigma_{k}^{+}+h.c,
\end{split}
\end{equation}
where the Lamb-Dicke matrix $\eta_{j,k}$ represents the participation of ion $k$ in motional mode $j$. The driving frequencies have been generalized to $\omega_{b,\pm}=\omega_0+\omega_b\pm\frac{n\Delta}{2}$. 

This also generalizes the definitions of $\alpha$ and $\beta$ with the modification $\xi_b\rightarrow\xi_{b,j}=\omega_b-\nu_j$. We assume that the limits $\alpha\gg1$ and $\beta\gg1$ hold regardless of the mode-index $j$. As such, the leading order contribution of the Magnus expansion, $\chi_1$ scales as $\left(\alpha\beta N\right)^{-1/2}$ and is therefore negligible.

We decompose the second order Magnus term to $\chi_2=\chi_{2,j}+\chi_{2,j,j^\prime}$. The former term does not couple between different normal modes of motion and therefore is a trivial generalization of Eq. \eqref{Om2}, it is given by,

\begin{equation}\label{Om2j}
\begin{split}
    \chi_{2,j}&=\chi_{2,j;\text{h}}+\chi_{2,j;z}\\
    \chi_{2,j;\text{h}}&=iT\hbar^{2}\Omega_{b}^{2}\sum_{k}B_{k,k+n}\sigma_{k+n}^{+}\sigma_{k}^{-}e^{i\phi}+h.c+\ord{\alpha^{-2}}\\
    \chi_{2,j;z}&=2iT\hbar^{2}\Omega_{b}^{2}\sum_{j,k}B_{k,k+n}\sigma_{k}^{z}\left(a_j^{\dag}a_j+\frac{1}{2}\right)+\ord{\alpha^{-2}}.
\end{split}
\end{equation}
where we defined $B_{i,k}=\sum_j \eta_{j,i}\eta_{j,k}/2\xi_{b,j}$. That is, we obtain the same effective Hamiltonians as in Eq. \eqref{Hterms_homo} but with the normalized spin-spin coupling $B_{i,k}$, which comes about due to contributions from all of the normal-modes, in analogy to \cite{porras2004effective}. 

Since the radial modes are bunched, typically $\left|\xi_{b,j}-\xi_{b,k}\right|\ll\left|\nu_j-\nu_k\right|$. Consequently the red sideband can be used in order to eliminate $H_z$, as described in the single normal-mode case above.

The term $\chi_{2,j,j^\prime}$ couples between mode $j$ and mode $j^\prime$. It only contains operators of the form $\sigma_{k}^{z}a_j^{\dag}a_{j^\prime}$, and its conjugate; therefore, this term generate no spin-hopping. By using the red sideband frequency pair, the leading order contribution of this term scales as $\alpha^{-2}$ and is therefore neligible in the large $\alpha$ limit.

We thus conclude that our method, in its continuous instance, is compatible with coupling through the radial motional modes, with the appropriate modification of coupling strengths.

\end{document}